\documentclass{article}

\usepackage[dblblindworkshop, final]{neurips_2026}
\workshoptitle{{SLMs for Agentic Systems, Paris, France, 2026}}

\usepackage[utf8]{inputenc} 
\usepackage[T1]{fontenc}    
\usepackage{hyperref}       
\usepackage{url}            
\usepackage{booktabs}       
\usepackage{amsfonts}       
\usepackage{nicefrac}       
\usepackage{microtype}      
\usepackage{xcolor}         

\usepackage{amsmath}
\usepackage{amssymb}
\usepackage{mathtools}
\usepackage{amsthm}

\usepackage{graphicx}
\usepackage{caption}
\usepackage{subcaption}
\usepackage{tabularx}
\usepackage{multirow}
\usepackage{float}
\usepackage{pifont}

\newcolumntype{Y}{>{\raggedleft\arraybackslash}X}

\usepackage{algorithm}
\usepackage{algpseudocode}

\usepackage{placeins}
\usepackage[capitalize,noabbrev]{cleveref}

\theoremstyle{plain}

\theoremstyle{definition}

\theoremstyle{remark}

\title{        KV-Pipe: On the Relation Between KV Sharing and
        Pipeline Parallel Efficiency in LLMs}

\author{%
  Maryam Dialameh \\
  Department of Mechanical and Mechatronics Engineering \\
  University of Waterloo \\
  Waterloo, Canada \\
  \texttt{maryam[dot]dialameh[at]uwaterloo[dot]ca}
  \And
  Hossein Rajabzadeh \\
  Department of Mechanical and Mechatronics Engineering \\
  University of Waterloo \\
  Waterloo, Canada
  \And
  Harish Krishnamoorthy Murali \\
  Ascend Team \\
  Huawei Technologies \\
  Toronto, Canada
  \And
  Walid Ahmed\thanks{This work was done while the author was affiliated with Huawei Canada.} \\
  Ascend Team \\
  Huawei Technologies \\
  Toronto, Canada
  \And
  Weiwei Zhang\footnotemark[1] \\
  Ascend Team \\
  Huawei Technologies \\
  Toronto, Canada
  \And
  Hyock Ju Kwon \\
  Department of Mechanical and Mechatronics Engineering \\
  University of Waterloo \\
  Waterloo, Canada
}

\begin{document}

\maketitle

\begin{abstract}
Pipeline parallelism (PP) is widely used to scale large language model (LLM)
training, but its efficiency is often limited by stage imbalance and pipeline
bubbles. Cross-layer KV sharing has primarily been studied for reducing KV-cache
costs during inference, without examining how KV reuse reshapes pipeline workloads.
We present \textbf{KV-Pipe}, a stage-aware KV-sharing mechanism that turns KV
reuse into a pipeline-balancing control knob. KV-Pipe starts from the tail stage,
converts selected attention layers to cross-layer KV sharing, and iteratively
retargets the current bottleneck to drive the FLOPs Imbalance Ratio (FIR) toward
$1$. The procedure runs offline using only the pipeline partition and per-layer
FLOP estimates, requiring no online tuning. Across multiple pipeline-parallel
configurations, KV-Pipe achieves up to \textbf{9.2\%} relative MFU improvement
and a \textbf{9.8\%} reduction in iteration time over the 1F1B baseline, with
larger gains at higher PP degrees. An inference case study further shows that
the same KV-sharing mechanism reduces KV-cache growth and redundant KV projection
work, improving long-context decoding throughput. These results identify KV
sharing as a system--architecture degree of freedom for jointly improving
pipeline-parallel training efficiency and long-context inference.
\end{abstract}

\section{Introduction}
Large language models (LLMs) now achieve state-of-the-art performance in complex reasoning tasks, mathematical problem solving, multi-step planning, and agentic behavior \cite{zhao2024docmath,toroghi2024verifiable,li2024gsm,schumann2024velma}. As model sizes continue to scale \cite{chowdhery2023palm,hoffmann2022training,smith2022using,adler2024nemotron}, both training and inference impose memory and compute demands far exceeding the capacity of a single accelerator.

\begin{figure}
    \centering
    \includegraphics[width=0.6\linewidth]{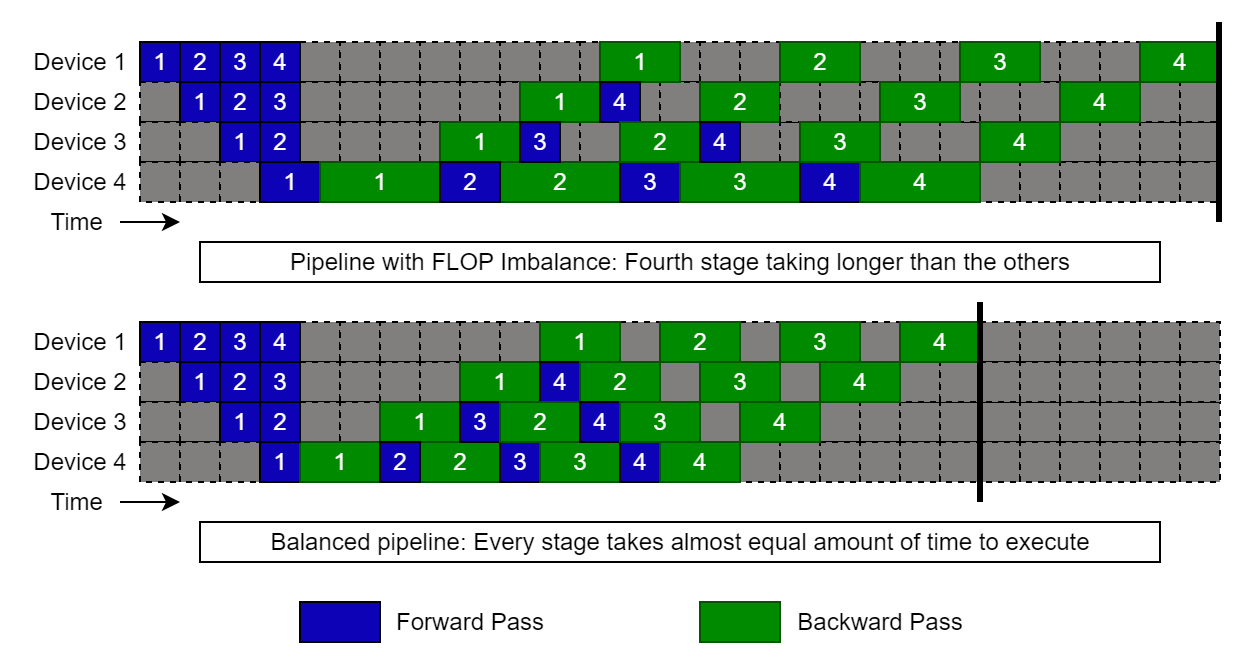}
        \caption{
        \textbf{Motivation.} 
        (a) Pipeline parallelism suffers from stage imbalance and bubble overhead. 
        (b) KV sharing restructures cross-layer attention workloads, enabling better FLOP distribution across stages. 
        (c) KV-Pipe leverages stage-aware KV sharing to reduce bubble time and improve MFU.
    }
    \label{fig:kvstrategies}
\end{figure}

Despite its attractive communication profile, \emph{pipeline parallelism} (PP) is notoriously sensitive to
\textit{stage imbalance} and \textit{pipeline bubbles}, which waste a non-trivial fraction of allocated accelerator time,
especially at scale \cite{huang2019gpipe,narayanan2019pipedream,narayanan2021efficient}.
Recent systems work has pushed PP efficiency by (i) \emph{re-partitioning} or \emph{rebalancing} layers to address memory and compute skew
(e.g., activation/memory balancing in BPipe and cost-model-driven partition search in DawnPiper) \cite{kim2023bpipe,peng2025dawnpiper},
(ii) \emph{mitigating bubbles} via scheduling or opportunistic backfilling (e.g., PipeFill fills bubble time with other jobs) \cite{arfeen2025pipefill},
and (iii) \emph{handling prefill--decode heterogeneity} in serving through phase-aware batching and throttling
(e.g., chunked-prefill scheduling in Sarathi/Sarathi-Serve, token throttling in gLLM, temporally-disaggregated execution in TD-Pipe,
and dynamic re-sharding in Seesaw) \cite{agrawal2023sarathi,agrawal2024taming,guo2025gllm,zhang2025td,su2025seesaw}.
These approaches substantially improve utilization, yet they largely treat the model’s per-layer cost profile as fixed, leaving a key
system-level degree of freedom underexplored: \emph{how the model’s internal KV layout interacts with pipeline partitioning}.

Modern LLMs increasingly depart from uniform Transformer stacks by combining heterogeneous blocks, including sliding-window attention \cite{jiang2023mistral7b}, sparse MoE layers \cite{jiang2024mixtralexperts}, and Transformer--Mamba hybrids \cite{lieber2024jambahybridtransformermambalanguage,blakeman2025nemotron}. This heterogeneity creates substantial variation in per-layer FLOPs, activation cost, and KV-cache growth, making pipeline stages difficult to balance. Cross-layer sharing offers one way to reduce this redundancy: ALBERT shares parameters across depth \cite{lan2019albert}, while recent KV-sharing methods such as MLKV and HShare reduce KV-cache memory and decoding overhead \cite{zuhri2025mlkv,wu2025hshare,wu2025systematic}. KV-cache management is itself a major serving bottleneck, as reflected in systems such as PagedAttention/vLLM \cite{kwon2023efficient}. However, existing KV-sharing work primarily targets memory or decoding efficiency rather than explicitly optimizing pipeline-stage balance.

This paper introduces \textbf{KV-Pipe}, a stage-aware KV-sharing framework that elevates KV reuse from a memory-compression trick to a pipeline-level control knob. The key novelty is to couple cross-layer KV sharing with PP partitioning: by converting carefully chosen attention layers (starting from the tail stage and then tracking the moving bottleneck), KV-Pipe reshapes the per-stage compute profile without changing the PP schedule, micro-batching, optimizer, or training dynamics. KV-Pipe diagnoses imbalance using a lightweight FLOPs Imbalance Ratio (FIR) and then applies \emph{bottleneck-targeted} KV sharing to directly flatten stage times (FIR$\!\to\!1$), reducing bubbles at their source rather than merely hiding them with scheduling. This yields a double dividend: during training, KV-Pipe increases PP utilization by reducing stage skew and idle time; during inference, the same KV sharing reduces KV-cache growth and redundant KV projections, enabling larger effective batching and higher long-context throughput. In effect, KV-Pipe exposes KV layout as a new, orthogonal and composable system--architecture degree of freedom that can be applied on top of existing PP methods to improve both train-time MFU and inference speed.

\section{Methodology} \label{sec:method} KV-Pipe improves pipeline-parallel (PP) efficiency by using cross-layer KV sharing \cite{dialameh2025echo,rajabzadeh2026efficient} as a \emph{stage-aware balancing knob}. Cross-layer KV sharing reduces the computation and KV-cache footprint of selected attention layers. Rather than applying sharing uniformly, KV-Pipe places these layers on expensive pipeline stages to reduce stage imbalance and pipeline bubbles. 

Consider a decoder-only Transformer with $L$ attention layers partitioned across $P$ pipeline stages, where $\mathcal{S}_i$ denotes the layers assigned to stage $i$. A standard attention layer independently computes its key and value projections, whereas a KV-sharing layer reuses them from an earlier layer $\tau(\ell)<\ell$: \begin{equation} \mathbf{K}_{\ell} \leftarrow \mathbf{K}_{\tau(\ell)}, \qquad \mathbf{V}_{\ell} \leftarrow \mathbf{V}_{\tau(\ell)}. \label{eq:kv_share} \end{equation} The query projection and attention computation remain unchanged. Consequently, converting layer $\ell$ to KV sharing removes redundant key/value projections and avoids storing an additional pair of KV states during autoregressive inference. We denote the corresponding FLOP reduction by $\Delta F_{\ell}=F_{\ell}(0)-F_{\ell}(1)>0$, where $F_{\ell}(0)$ and $F_{\ell}(1)$ are the costs of the full-attention and KV-sharing variants, respectively. Let $z_{\ell}\in\{0,1\}$ indicate whether layer $\ell$ uses KV sharing. The FLOPs assigned to stage $i$ are \begin{equation} F_i(\mathbf{z}) = \sum_{\ell\in\mathcal{S}_i} F_{\ell}(z_{\ell}) + C_i, \label{eq:stage_flops_z} \end{equation} where $C_i$ captures stage-specific architectural costs, such as the embedding module or LM head. These non-uniform components can make a nominally uniform layer partition substantially imbalanced. To quantify this imbalance, we introduce the \textbf{FLOPs Imbalance Ratio (FIR)}: \begin{equation} \mathrm{FIR}(\mathbf{z}) = \frac{\max_i F_i(\mathbf{z})} {\frac{1}{P}\sum_{i=1}^{P}F_i(\mathbf{z})}. \label{eq:flop_imbalance_main} \end{equation} A perfectly balanced pipeline has $\mathrm{FIR}=1$, while larger values indicate that one or more stages contain disproportionately more computation. Because pipeline throughput is constrained by the slowest stage, reducing FIR provides a simple architecture-level proxy for reducing pipeline bubbles. 

\subsection{Architecture-Balanced KV Placement} \label{sec:kvpipe} Given a fixed PP partition and a budget of at most $m$ KV-sharing layers, KV-Pipe seeks a placement that reduces the bottleneck stage: \begin{equation} \min_{\mathbf{z}}\; \mathrm{FIR}(\mathbf{z}) \qquad \text{s.t.}\qquad \sum_{\ell=1}^{L} z_{\ell}\le m. \label{eq:kvpipe_objective} \end{equation} Rather than solving this discrete optimization exactly, we use a lightweight greedy policy. KV-Pipe first targets the tail stage, which is frequently the initial bottleneck because it also contains the LM head. Within that stage, layers are converted in a last-layer-first order. After every conversion, stage FLOPs and FIR are updated. Once the tail stage is no longer the bottleneck, KV-Pipe continues from the current maximum-FLOP stage. The procedure stops when $\mathrm{FIR}\le 1+\epsilon$, the conversion budget is exhausted, or no eligible layer remains. This policy directly spends the KV-sharing budget where it provides the largest benefit to pipeline balance, while simultaneously reducing KV-cache consumption. It is also inexpensive: KV-Pipe requires only the PP partition and per-layer FLOP estimates, obtained analytically or from a one-time profiling pass, and runs offline before training or inference. We additionally study Uniform and Symmetric-Bipolar placement policies to isolate the effect of stage-aware placement. Appendix~\ref{app:additional_analysis} provides their definitions, an illustrative PP partition, the complete KV-Pipe algorithm, and a case study showing how the different policies affect FIR.

\paragraph{KV source selection.}
In all experiments, a KV-sharing layer reuses the key/value (K/V) states of the immediately preceding layer, i.e.,
$\tau(\ell)=\ell-1$.
For consecutive KV-sharing layers, this reuse is chained: a sequence of shared layers
$\ell,\ldots,\ell+r$ ultimately reuses the K/V states produced by the nearest preceding full-attention layer.
For example, when layers 29--32 of LLaMA2-7B are converted to KV-sharing layers, their K/V states originate from layer 28.

\section{Experiments}
\label{sec:exp}

We evaluate KV-Pipe on \textbf{LLaMA2-7B}  as our primary setup, and further validate it on \textbf{LLaMA2-7B/13B}, \textbf{LLaMA3-8B}, and \textbf{Qwen2.5-14B}. We cover context lengths from \textbf{4K to 16K}; additional long-context inference and prior-method comparisons are reported in Appendix~\ref{app:additional_analysis}. Unless otherwise stated, training uses fixed global batch size and optimizer/precision settings. All experiments use pipeline parallelism with degree $\text{PP}\in\{2,4,8\}$. We keep other parallelism dimensions fixed (i.e., we do not set or vary TP/CP in the main experiments) to isolate PP effects.

We report model FLOPs utilization (MFU) as
\begin{equation}
\mathrm{MFU}
=
\frac{F_{\text{model}}\,T_{\text{iter}}}{F_{\text{peak}}},
\label{eq:mfu_def}
\end{equation}
where $F_{\text{model}}$ is the model compute per iteration (FLOPs/iter),
$T_{\text{iter}}$ is the measured iteration throughput (iter/s), and
$F_{\text{peak}}$ is the device peak compute (FLOPs/s), scaled by the number of devices used. The immediately preceding layer provides key-value states for the KV-sharing layers. We compare three KV-sharing placement strategies under PP:
\textsc{Uniform}, \textsc{Symmetric Bipolar}, and \textsc{Architecture-Balanced}.
We also vary the KV-sharing budget (number of KV-sharing layers) for ablation.

\paragraph{Shared-KV model preparation.}
The shared-KV models are not trained from scratch. We initialize from the
corresponding released pretrained LLaMA checkpoints, convert the selected
attention layers to the cross-layer KV-sharing architecture, and then perform
continual training to adapt the pretrained model to the modified attention
structure. We follow the continual-training data mixture and optimization
configuration of ECHO-LLaMA~\cite{dialameh2025echo}. Unless otherwise stated, the same adaptation
protocol is used across the evaluated KV-sharing configurations; only the
number and placement of shared-KV layers are varied.


\textbf{Results on Ascend 910B NPUs.}
\label{sec:exp_npu_results}

\begin{figure*}[t]
    \centering
    \begin{subfigure}[t]{0.32\textwidth}
        \centering
        \includegraphics[width=\linewidth]{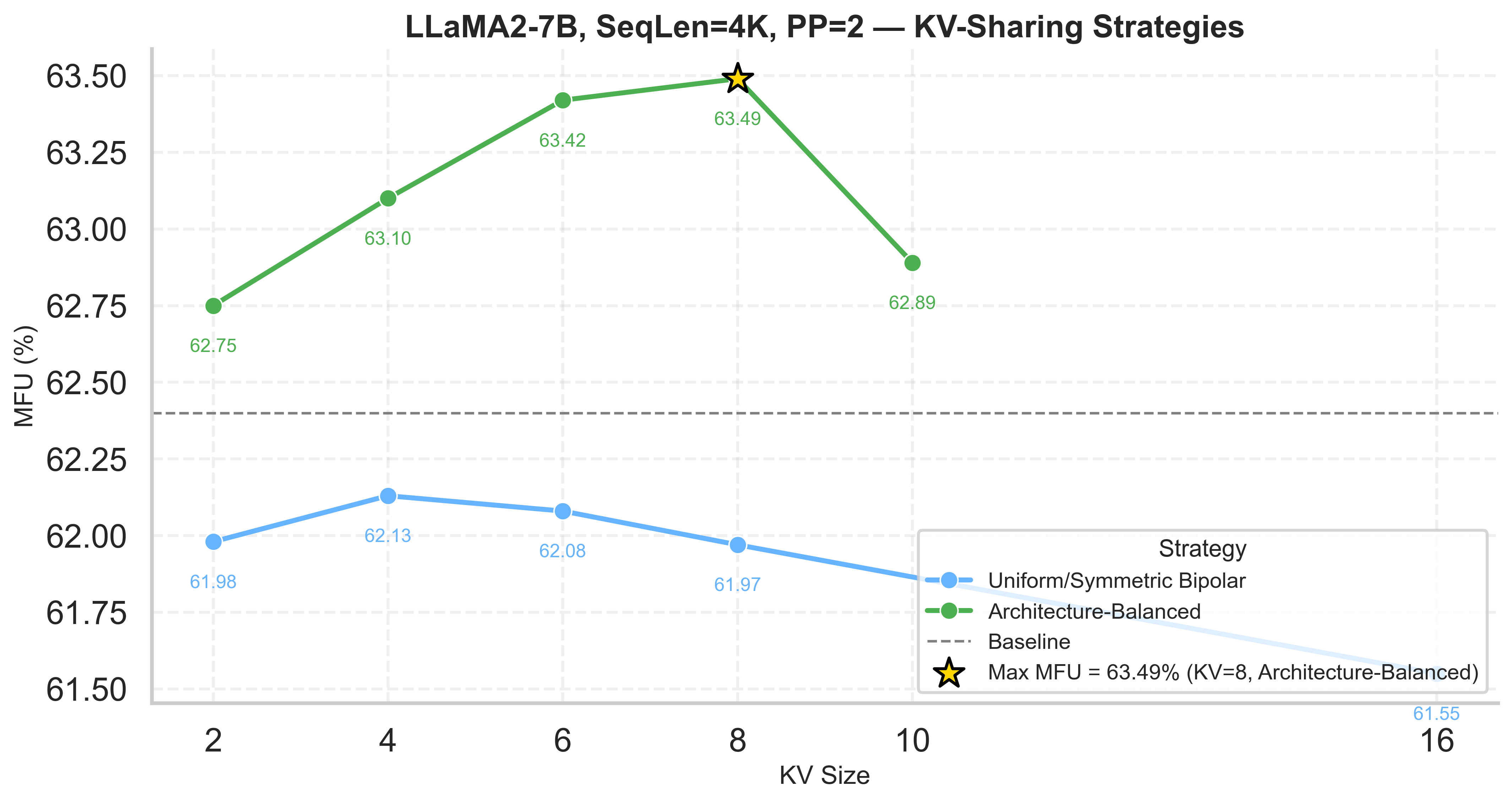}
        \caption{$S{=}4$K, PP=2}
    \end{subfigure}\hfill
    \begin{subfigure}[t]{0.32\textwidth}
        \centering
        \includegraphics[width=\linewidth]{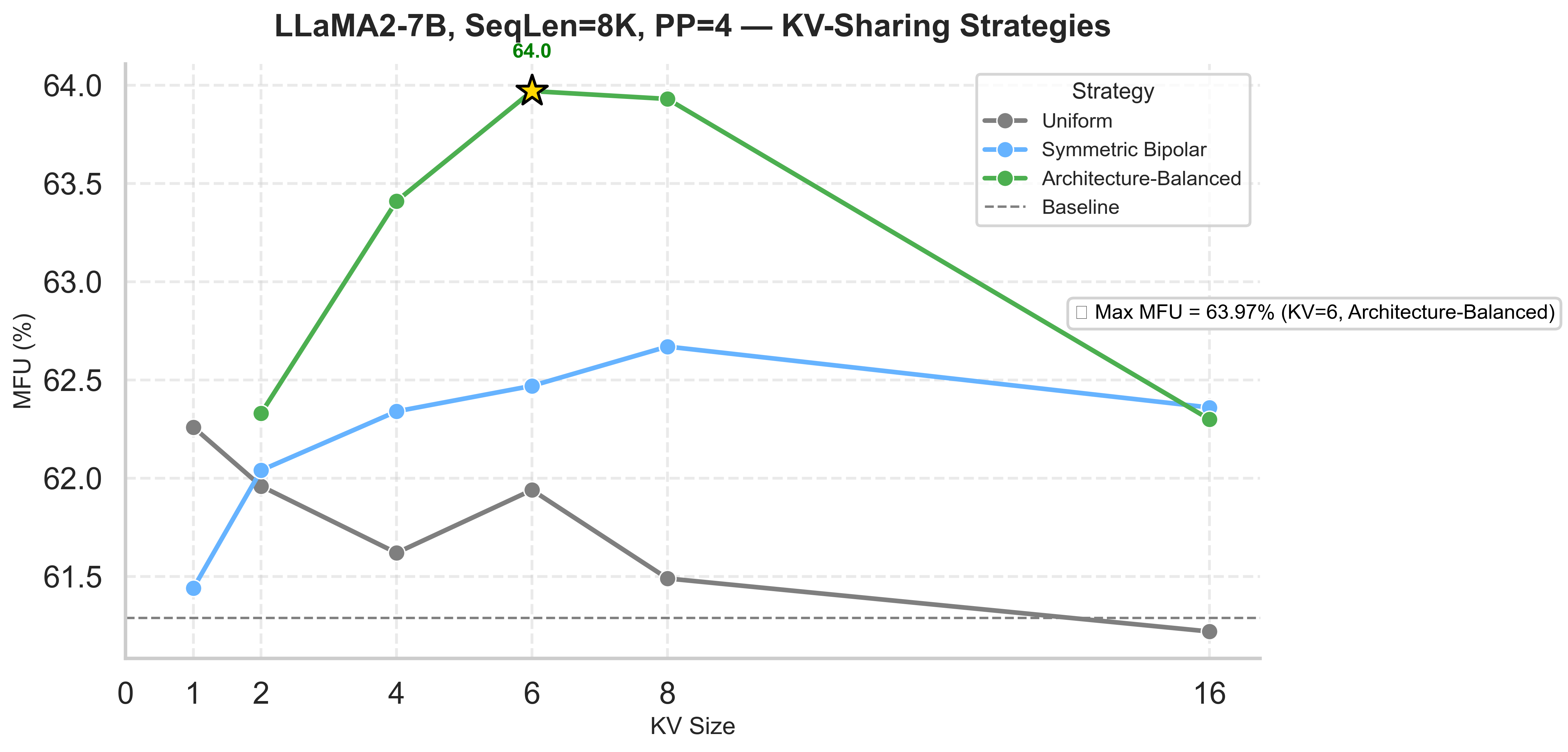}
        \caption{$S{=}8$K, PP=4}
    \end{subfigure}\hfill
    \begin{subfigure}[t]{0.32\textwidth}
        \centering
        \includegraphics[width=\linewidth]{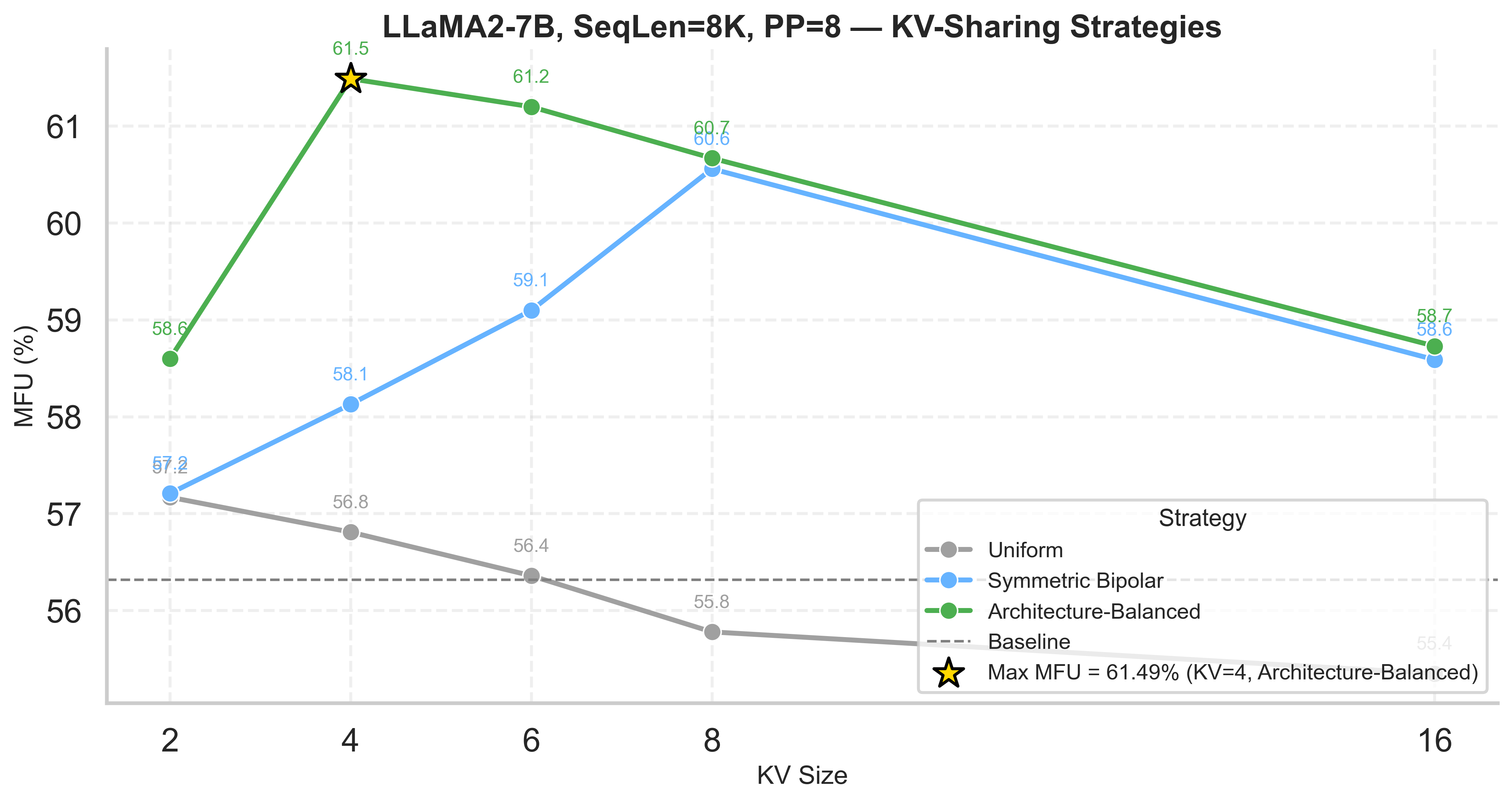}
        \caption{$S{=}8$K, PP=8}
    \end{subfigure}
    \caption{\textbf{MFU sweeps on 8$\times$ Ascend 910B.}
MFU vs.\ KV-sharing budget for the three KV-Pipe placements. Stars indicate the best MFU point used in Table~\ref{tab:npu_best_subtables}.}
    \label{fig:npu_mfu_sweeps}
\end{figure*}

\begin{figure*}[t]
    \centering
    \begin{subfigure}[t]{0.32\textwidth}
        \centering
        \includegraphics[width=\linewidth]{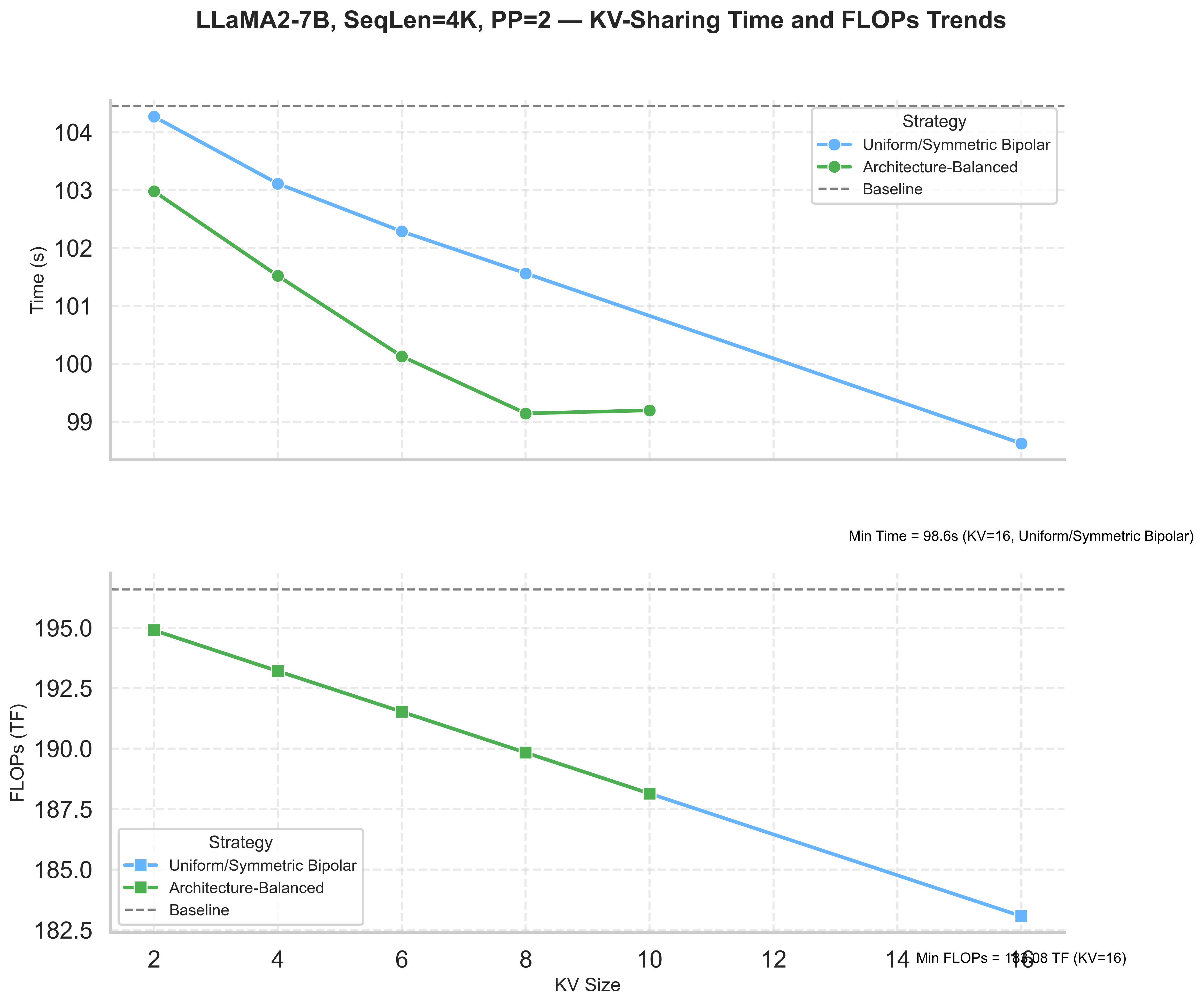}
        \caption{$S{=}4$K, PP=2}
    \end{subfigure}\hfill
    \begin{subfigure}[t]{0.32\textwidth}
        \centering
        \includegraphics[width=\linewidth]{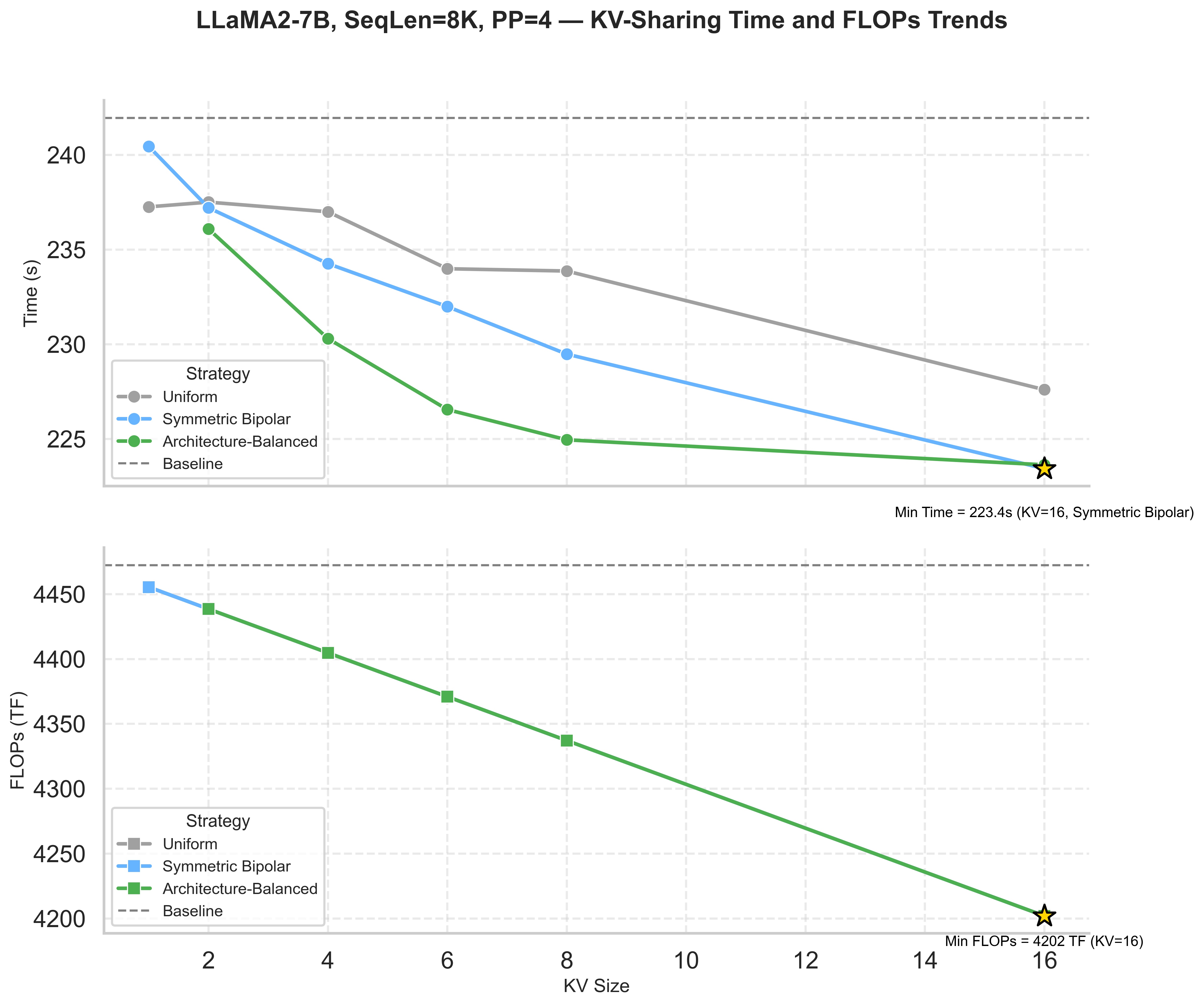}
        \caption{$S{=}8$K, PP=4}
    \end{subfigure}\hfill
    \begin{subfigure}[t]{0.32\textwidth}
        \centering
        \includegraphics[width=\linewidth]{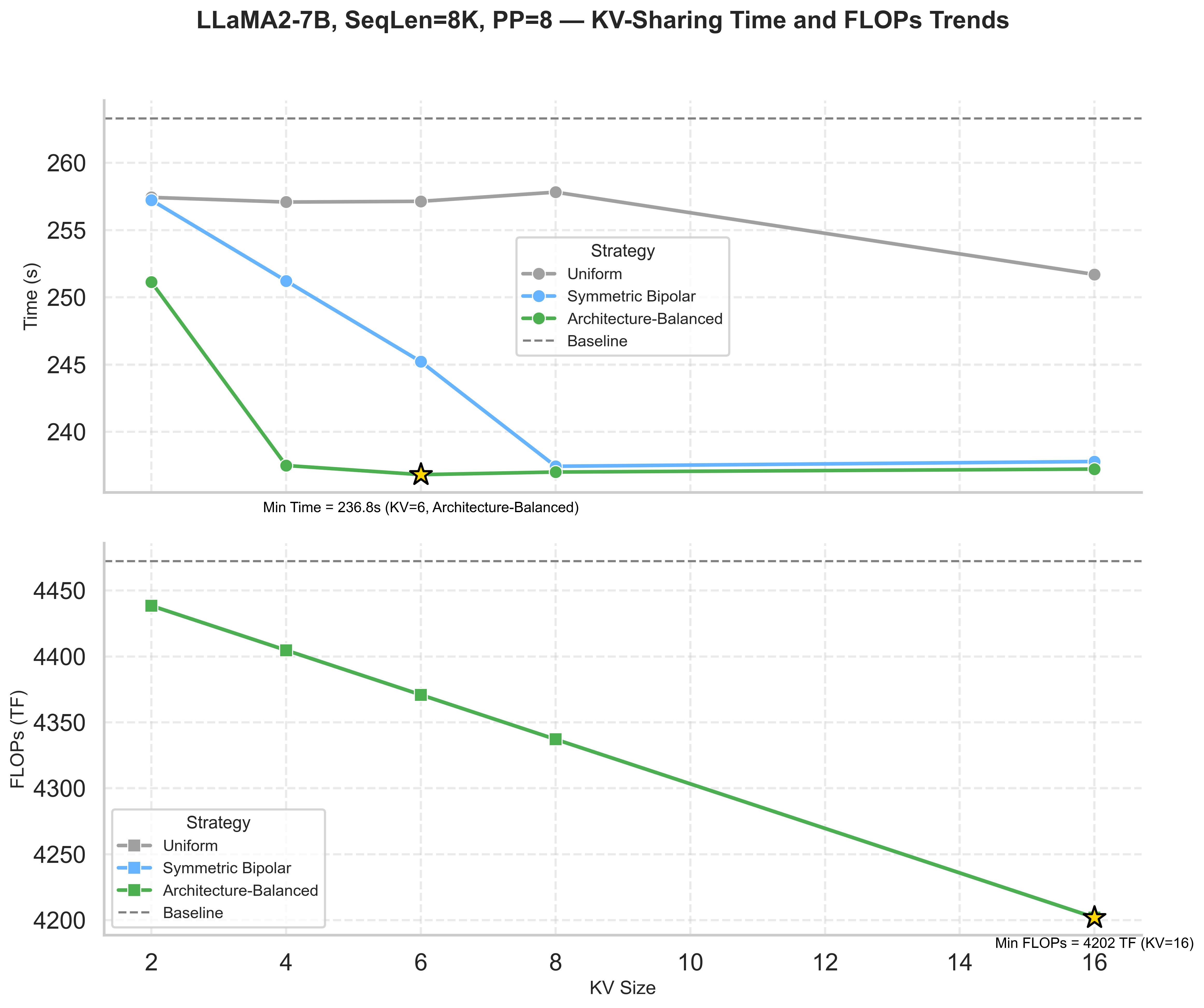}
        \caption{$S{=}8$K, PP=8}
    \end{subfigure}
    \caption{\textbf{Iteration time and FLOPs sweeps on 8$\times$ Ascend 910B.}
Each panel reports iteration time (top) and total model FLOPs (bottom) as KV-sharing budget increases, for all placements.}
    \label{fig:npu_timeflops_sweeps}
\end{figure*}

We evaluate KV-Pipe on \textbf{8$\times$ Ascend 910B NPUs} using MindSpeed-LLM’s \textbf{1F1B} pipeline schedule (no interleaving)
with \textbf{LLaMA2-7B}. We test three training configurations:
(1) $S{=}4$K, PP=2; (2) $S{=}8$K, PP=4; (3) $S{=}8$K, PP=8.
For each configuration, we sweep the KV-sharing budget (``KV size'') under the three placement strategies
and select the \textbf{best KV-Pipe variant}, which is consistently the \textbf{Architecture-Balanced} placement.
Table~\ref{tab:npu_best_subtables} summarizes the \emph{baseline 1F1B (no KV sharing)} versus the \emph{best KV-Pipe (Architecture-Balanced)}
at the MFU-optimal KV-sharing budget for each configuration; the full MFU sweeps and time/FLOPs sweeps are shown in
Figures~\ref{fig:npu_mfu_sweeps} and~\ref{fig:npu_timeflops_sweeps}.

\textbf{MFU and time improvements.}
Across all settings, KV-Pipe increases MFU and reduces iteration time.
The MFU gains grow with pipeline depth: from 62.45$\rightarrow$64.39 (PP=2) to 61.28$\rightarrow$63.93 (PP=4),
and 56.32$\rightarrow$61.49 (PP=8), corresponding to relative MFU improvements of 3.10\%, 4.32\%, and 9.17\%, respectively
(Table~\ref{tab:npu_best_subtables}). Iteration time decreases by 4.90\%--9.80\%, indicating consistent end-to-end throughput gains.

\textbf{Imbalance metric tracks utilization.}
The best KV-Pipe setting also moves FIR closer to 1 in all three configurations
(e.g., 1.1194$\rightarrow$1.0004 for $S{=}8$K, PP=8), matching the trend in MFU improvements and supporting our hypothesis
that reducing stage imbalance is a key driver of PP utilization (Table~\ref{tab:npu_best_subtables}).Figures~\ref{fig:npu_mfu_sweeps} and~\ref{fig:npu_timeflops_sweeps} reveal a consistent \emph{non-monotonic} MFU pattern as we increase the
KV-sharing budget (shared-KV size). At small shared-KV sizes, converting a few full-attention layers to KV-sharing layers in the \textsc{Architecture-Balanced}
placement selectively reduces the \emph{bottleneck stage} compute, thereby reducing the stage-time skew. Equivalently, the Architecture-Balanced strategy reduces FIR and moves it closer to the ideal value of 1 (Table~\ref{tab:npu_best_subtables}), indicating a more balanced pipeline.
This initial balancing effect increases MFU sharply (e.g., the green curve in Figure~\ref{fig:npu_mfu_sweeps}), because less time is wasted in pipeline bubbles and idle periods. Importantly, the KV-sharing size that maximizes MFU is an \emph{intermediate} point rather than the largest shared-KV size.
Across our three NPU settings, the MFU-optimal shared-KV sizes are KV=8 for ($S{=}4$K, PP=2), KV=6 for ($S{=}8$K, PP=4), and KV=4 for ($S{=}8$K, PP=8),
as highlighted by stars in Figure~\ref{fig:npu_mfu_sweeps} and summarized in Table~\ref{tab:npu_best_subtables}. These optima occur where KV-Pipe has
sufficiently reduced the bottleneck stage to \emph{equalize stage latencies}, i.e., FIR is closest to 1, so the pipeline achieves near-maximal overlap.

\textbf{Why MFU drops beyond the optimum: \emph{over-correction} induces new imbalance.}
When shared-KV size increases beyond the MFU-optimal point, KV-Pipe begins to \emph{over-lighten} the previously bottlenecked stages.
While total model FLOPs continue to decrease monotonically (bottom rows of Figure~\ref{fig:npu_timeflops_sweeps}), the \emph{per-stage} compute distribution
is no longer aligned: the earlier (or middle) stages can become the new bottleneck, and FIR increases again (moving away from 1). This renewed imbalance
reintroduces pipeline idle time, and MFU consequently declines after the peak (Figure~\ref{fig:npu_mfu_sweeps}). This explains why ``more KV sharing'' is not
always better for PP efficiency: the dominant factor is \emph{balance}, not just fewer FLOPs.

\textbf{Time/FLOPs sweeps corroborate the MFU behavior.}
Figure~\ref{fig:npu_timeflops_sweeps} further supports this interpretation. FLOPs decrease approximately monotonically with KV size for all strategies,
yet iteration time shows diminishing returns and can flatten (or even slightly worsen) once imbalance shifts to other stages. In practice, the shared-KV size that
minimizes time does not necessarily coincide with the shared-KV size that maximizes MFU, because MFU is sensitive to bubble overhead caused by imbalance.
Overall, these results validate the design principle of KV-Pipe: \emph{stage-aware KV placement should be tuned to minimize imbalance (FIR$\rightarrow$1),}
which yields an optimal KV-sharing budget that maximizes PP efficiency.

\begin{table*}[t]
\centering

\begin{subtable}[t]{0.32\textwidth}
\centering
\caption{$S{=}4$K, PP=2, SKV=8}
\label{tab:npu_4k_pp2}

\scriptsize
\setlength{\tabcolsep}{2pt}
\renewcommand{\arraystretch}{1.05}

\begin{tabular}{@{}
    >{\raggedright\arraybackslash}p{0.31\linewidth}
    >{\centering\arraybackslash}p{0.43\linewidth}
    >{\centering\arraybackslash}p{0.18\linewidth}
@{}}
\toprule
\textbf{Metric} & \textbf{Change} & \textbf{Gain} \\
\midrule
MFU
& 62.45$\rightarrow$\textbf{64.39}
& 3.10\% \\

Iteration time (s)
& 104.14$\rightarrow$\textbf{99.27}
& 4.90\% \\

FIR
& 1.0182$\rightarrow$\textbf{1.0101}
& 0.80\% \\

FWD time
& 0.1152$\rightarrow$\textbf{0.1094}
& 5.30\% \\
\bottomrule
\end{tabular}
\end{subtable}
\hfill
%
\begin{subtable}[t]{0.32\textwidth}
\centering
\caption{$S{=}8$K, PP=4, SKV=6}
\label{tab:npu_8k_pp4}

\scriptsize
\setlength{\tabcolsep}{2pt}
\renewcommand{\arraystretch}{1.05}

\begin{tabular}{@{}
    >{\raggedright\arraybackslash}p{0.31\linewidth}
    >{\centering\arraybackslash}p{0.43\linewidth}
    >{\centering\arraybackslash}p{0.18\linewidth}
@{}}
\toprule
\textbf{Metric} & \textbf{Change} & \textbf{Gain} \\
\midrule
MFU
& 61.28$\rightarrow$\textbf{63.93}
& 4.32\% \\

Iteration time (s)
& 241.96$\rightarrow$\textbf{224.95}
& 7.56\% \\

FIR
& 1.051$\rightarrow$\textbf{1.018}
& 3.24\% \\

FWD time
& 0.1260$\rightarrow$\textbf{0.1189}
& 5.97\% \\
\bottomrule
\end{tabular}
\end{subtable}
\hfill
%
\begin{subtable}[t]{0.32\textwidth}
\centering
\caption{$S{=}8$K, PP=8, SKV=4}
\label{tab:npu_8k_pp8}

\scriptsize
\setlength{\tabcolsep}{2pt}
\renewcommand{\arraystretch}{1.05}

\begin{tabular}{@{}
    >{\raggedright\arraybackslash}p{0.31\linewidth}
    >{\centering\arraybackslash}p{0.43\linewidth}
    >{\centering\arraybackslash}p{0.18\linewidth}
@{}}
\toprule
\textbf{Metric} & \textbf{Change} & \textbf{Gain} \\
\midrule
MFU
& 56.32$\rightarrow$\textbf{61.49}
& 9.17\% \\

Iteration time (s)
& 263.3$\rightarrow$\textbf{237.5}
& 9.80\% \\

FIR
& 1.1194$\rightarrow$\textbf{1.0004}
& 11.89\% \\

FWD time
& 0.0347$\rightarrow$\textbf{0.0302}
& 14.90\% \\
\bottomrule
\end{tabular}
\end{subtable}

\caption{\textbf{NPU best-result summary (baseline vs.\ KV-Pipe).}
Each subtable compares baseline 1F1B (no KV sharing) against the best KV-Pipe
configuration (Architecture-Balanced placement at the MFU-optimal KV-sharing
budget) for the corresponding PP/sequence-length setting.
All subtables use LLaMA2-7B as the underlying LLM.
Here, $S$ denotes sequence length and $\mathrm{SKV}$ denotes shared-KV size.}
\label{tab:npu_best_subtables}

\end{table*}


\textbf{Results on GPUs (8$\times$V100).}
\label{sec:exp_gpu}
We repeat the PP-focused evaluation on a single node with \textbf{8$\times$NVIDIA V100} GPUs. We compare
\textbf{1F1B} \cite{harlap1806pipedream}, \textbf{Seq1F1B} \cite{ao2025seq1f1b}, and \textbf{KV-Pipe} under PP$\in\{2,4,8\}$ and context length
$S\in\{4\text{K},8\text{K},16\text{K}\}$, on four LLMs: LLaMA2-7B, LLaMA2-13B, LLaMA3-8B, and Qwen2.5-14B. Table~\ref{tab:gpu_pp_results_compact} reports PP-only performance. Each cell contains a triplet “PP$=$2 / 4 / 8”, i.e., the first number corresponds to PP=2, the second to PP=4, and the third to PP=8. Across all models and sequence lengths, \textbf{Seq1F1B} improves over \textbf{1F1B} by reducing pipeline bubbles through sequential scheduling, which is
most visible as PP increases (e.g., LLaMA2-7B at $S{=}8$K: MFU 39.0$\rightarrow$47.0 and time 305$\rightarrow$265 when moving from 1F1B to Seq1F1B at PP=8).
\textbf{KV-Pipe} provides an additional gain beyond Seq1F1B by directly reducing stage-level compute skew via stage-aware KV sharing, leading to both higher MFU and lower iteration time across the board. The benefit grows with pipeline depth: KV-Pipe’s improvements are modest at PP=2 but become consistently
largest at PP=8, where imbalance and idle time are amplified (e.g., LLaMA2-13B at $S{=}16$K and PP=8: MFU 39.5$\rightarrow$52.0 and time 960$\rightarrow$750 from 1F1B to KV-Pipe).
Longer contexts further strengthen the trend because KV/attention costs become more dominant and heterogeneous across stages, increasing the headroom for KV-Pipe
to rebalance the pipeline. Overall, the table shows that Seq1F1B primarily hides bubbles through better scheduling, while KV-Pipe removes bubbles by
making stage workloads more uniform; the combined effect is most pronounced at high PP and long context lengths.
\begin{table*}[t]
\centering

\caption{\textbf{GPU results on 8$\times$V100 (compact).}
Each entry reports PP=2/4/8 in a single cell.
KV-Pipe consistently improves MFU and reduces iteration time,
with larger gains at PP=8.}
\label{tab:gpu_pp_results_compact}

\small
\setlength{\tabcolsep}{3pt}
\renewcommand{\arraystretch}{1.05}

\resizebox{\textwidth}{!}{%
\begin{tabular}{@{}llcccccc@{}}
\toprule

\textbf{Model} &
\textbf{$S$} &
\multicolumn{3}{c}{\textbf{MFU (\%) at PP=2 / 4 / 8}} &
\multicolumn{3}{c}{\textbf{Iter. time (s) at PP=2 / 4 / 8}} \\

\cmidrule(lr){3-5}
\cmidrule(lr){6-8}

& &
\textbf{1F1B} &
\textbf{Seq1F1B} &
\textbf{KV-Pipe} &
\textbf{1F1B} &
\textbf{Seq1F1B} &
\textbf{KV-Pipe} \\

\midrule

LLaMA2-7B
& 4K
& 46.5/44.0/38.0
& 49.5/48.0/45.0
& 51.5/50.0/48.0
& 112/125/150
& 106/114/128
& 101/109/120 \\

& 8K
& 48.0/46.0/39.0
& 50.0/49.5/47.0
& 52.0/52.0/50.0
& 215/265/305
& 205/245/265
& 196/232/248 \\

& 16K
& 49.0/46.5/40.0
& 51.0/50.0/48.0
& 53.0/52.5/51.0
& 420/520/610
& 400/470/520
& 385/448/485 \\

\midrule

LLaMA2-13B
& 4K
& 47.0/44.5/38.5
& 49.0/48.5/46.0
& 51.0/50.5/49.0
& 195/220/260
& 186/204/228
& 176/192/212 \\

& 8K
& 48.5/45.0/39.0
& 50.0/49.5/47.5
& 52.0/52.5/50.5
& 365/430/505
& 350/398/445
& 334/372/410 \\

& 16K
& 49.5/45.5/39.5
& 51.0/50.0/48.5
& 53.0/53.0/52.0
& 705/820/960
& 675/740/820
& 640/690/750 \\

\midrule

LLaMA3-8B
& 4K
& 46.0/43.5/37.5
& 49.0/47.8/45.0
& 51.0/49.8/48.0
& 118/132/158
& 111/121/136
& 105/114/126 \\

& 8K
& 47.5/45.0/38.5
& 49.5/49.0/46.8
& 51.5/51.8/49.8
& 228/278/320
& 217/257/278
& 205/240/258 \\

& 16K
& 48.5/45.2/39.0
& 50.5/49.6/47.8
& 52.5/52.0/51.0
& 445/545/635
& 420/495/545
& 402/468/505 \\

\midrule

Qwen2.5-14B
& 4K
& 47.5/45.0/39.0
& 49.8/49.5/47.0
& 52.0/51.8/50.0
& 210/236/278
& 200/218/242
& 188/202/222 \\

& 8K
& 49.0/45.5/39.5
& 50.5/50.2/48.0
& 52.5/53.0/51.0
& 392/465/540
& 375/428/472
& 356/398/430 \\

& 16K
& 50.0/46.0/40.0
& 51.5/51.0/49.0
& 53.5/54.0/52.5
& 760/885/1030
& 725/800/880
& 680/735/800 \\

\bottomrule
\end{tabular}%
}

\end{table*}
Across all evaluated models, the benefit of Seq1F1B grows with $P$ (bubble mitigation), while KV-Pipe yields an additional improvement by directly reducing stage-level skew; the combined effect is most pronounced at PP=8 and longer contexts where both bubble overhead and per-stage KV/compute heterogeneity are amplified.

\begin{figure}[!b]
    \centering
    \includegraphics[width=0.4\linewidth]{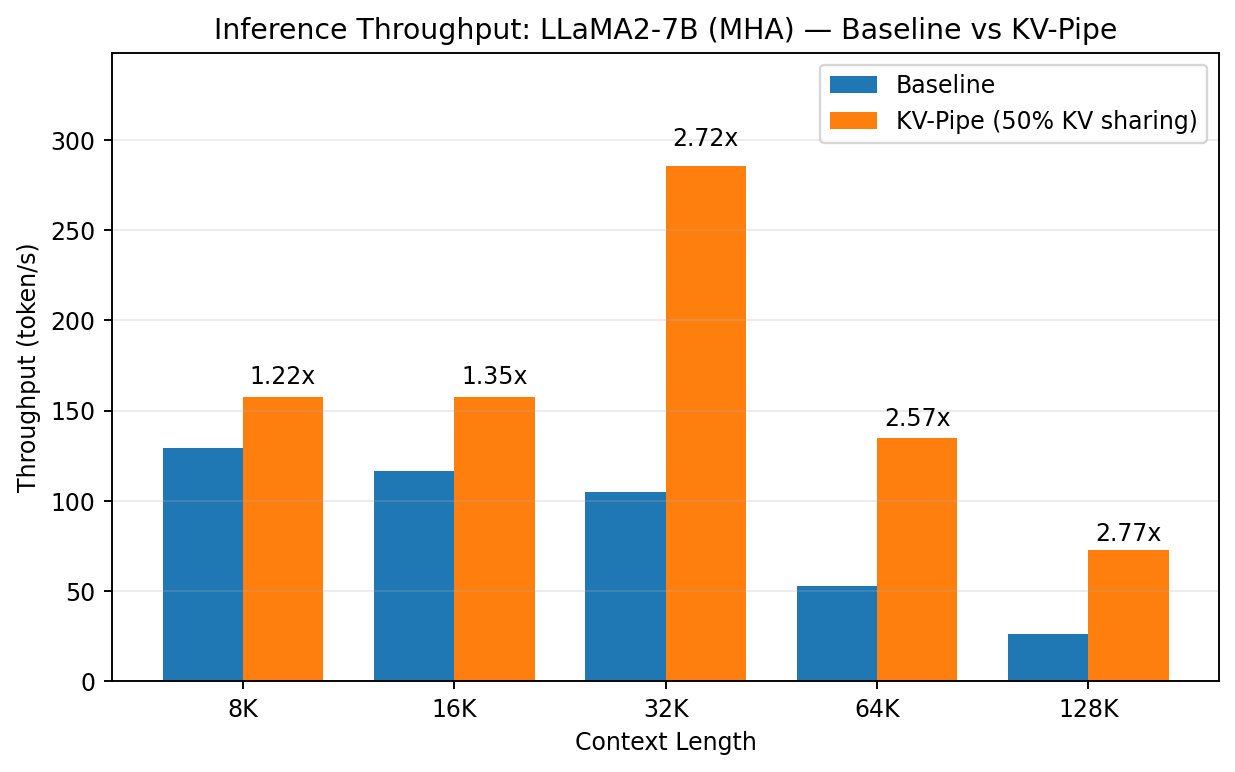}
    \caption{\textbf{Inference throughput (LLaMA2-7B, MHA).} Tokens/s vs.\ context length for the Transformer baseline and KV-Pipe with $50\%$ KV sharing, i.e. shared-KV size.}
    \label{fig:kvpipe_infer_llama2_7b_mha}
\end{figure}

\textbf{Inference case study (MHA) and double benefit of KV-Pipe.}
To complement our training results, we run an \emph{inference} case study to illustrate that KV-Pipe provides a second, orthogonal benefit beyond pipeline balancing.
We consider \textbf{LLaMA2-7B} with \textbf{standard Multi-Head Attention (MHA)} (i.e., each layer maintains its own $K,V$ cache per head, without grouped-query sharing),
and evaluate long-context decoding at sequence lengths $S\in\{8\text{K},16\text{K},32\text{K},64\text{K},128\text{K}\}$.
Figure~\ref{fig:kvpipe_infer_llama2_7b_mha} reports the end-to-end inference throughput (tokens/s) comparing a \textbf{Transformer baseline} (no KV sharing)
against \textbf{KV-Pipe} under identical execution assumptions. We observe a consistent throughput improvement that becomes
more pronounced as context grows: KV-Pipe achieves modest gains at shorter contexts (where cache pressure is limited), but delivers substantially larger speedups
at long contexts where KV-cache memory and attention cost dominate. Importantly, this inference improvement is additive to KV-Pipe’s training-side gains:
during training, KV-Pipe increases MFU by reducing stage-level compute skew and pipeline bubbles; during inference, it reduces KV-cache footprint and redundant KV
projection work, enabling higher effective throughput. This double benefit highlights KV-Pipe as a \textbf{generic and native system-level knob}:
it simultaneously (i) balances pipeline stages to improve training efficiency under PP and (ii) accelerates inference via cross-layer KV sharing,
making KV layout an actionable degree of freedom for optimizing both train-time MFU and inference-time speed.

\section{Conclusion}
\label{sec:conclusion}
This paper identifies an underexplored lever: the interaction between KV layout and pipeline partitioning. We introduce \textbf{KV-Pipe}, a stage-aware cross-layer KV-sharing framework that uses a simple FLOPs Imbalance Ratio (FIR) to locate bottleneck stages and then applies targeted KV sharing to flatten stage workloads (FIR$\!\to\!1$) while simultaneously reducing KV-cache growth. On 8$\times$Ascend 910B across PP$\in\{2,4,8\}$, KV-Pipe consistently improves end-to-end utilization, delivering \textbf{3.1--9.2\%} MFU gains and \textbf{4.9--9.8\%} iteration-time reductions over baseline 1F1B. We also observe a clear operating principle: MFU peaks at an \emph{intermediate} shared-KV budget—where imbalance is minimized—and can decline when excessive sharing over-corrects the original bottleneck and shifts the critical stage elsewhere. Overall, KV-Pipe establishes KV sharing as a system-level control knob that is orthogonal to PP scheduling: it directly reduces stage skew for higher train-time MFU and, by lowering KV-cache pressure, also improves long-context inference throughput.


\newpage
\bibliographystyle{plainnat}
\bibliography{ref}

@inproceedings{zhao2024docmath,
  title={DocMath-eval: Evaluating math reasoning capabilities of LLMs in understanding long and specialized documents},
  author={Zhao, Yilun and Long, Yitao and Liu, Hongjun and Kamoi, Ryo and Nan, Linyong and Chen, Lyuhao and Liu, Yixin and Tang, Xiangru and Zhang, Rui and Cohan, Arman},
  booktitle={Proceedings of the 62nd Annual Meeting of the Association for Computational Linguistics (Volume 1: Long Papers)},
  pages={16103--16120},
  year={2024}
}

@inproceedings{toroghi2024verifiable,
  title={Verifiable, debuggable, and repairable commonsense logical reasoning via llm-based theory resolution},
  author={Toroghi, Armin and Guo, Willis and Pesaranghader, Ali and Sanner, Scott},
  booktitle={Proceedings of the 2024 Conference on Empirical Methods in Natural Language Processing},
  pages={6634--6652},
  year={2024}
}

@article{li2024gsm,
  title={Gsm-plus: A comprehensive benchmark for evaluating the robustness of llms as mathematical problem solvers},
  author={Li, Qintong and Cui, Leyang and Zhao, Xueliang and Kong, Lingpeng and Bi, Wei},
  journal={arXiv preprint arXiv:2402.19255},
  year={2024}
}

@inproceedings{schumann2024velma,
  title={Velma: Verbalization embodiment of llm agents for vision and language navigation in street view},
  author={Schumann, Raphael and Zhu, Wanrong and Feng, Weixi and Fu, Tsu-Jui and Riezler, Stefan and Wang, William Yang},
  booktitle={Proceedings of the AAAI Conference on Artificial Intelligence},
  volume={38},
  number={17},
  pages={18924--18933},
  year={2024}
}

@article{chowdhery2023palm,
  title={Palm: Scaling language modeling with pathways},
  author={Chowdhery, Aakanksha and Narang, Sharan and Devlin, Jacob and Bosma, Maarten and Mishra, Gaurav and Roberts, Adam and Barham, Paul and Chung, Hyung Won and Sutton, Charles and Gehrmann, Sebastian and others},
  journal={Journal of Machine Learning Research},
  volume={24},
  number={240},
  pages={1--113},
  year={2023}
}

@article{hoffmann2022training,
  title={Training compute-optimal large language models},
  author={Hoffmann, Jordan and Borgeaud, Sebastian and Mensch, Arthur and Buchatskaya, Elena and Cai, Trevor and Rutherford, Eliza and Casas, Diego de Las and Hendricks, Lisa Anne and Welbl, Johannes and Clark, Aidan and others},
  journal={arXiv preprint arXiv:2203.15556},
  year={2022}
}

@article{smith2022using,
  title={Using deepspeed and megatron to train megatron-turing nlg 530b, a large-scale generative language model},
  author={Smith, Shaden and Patwary, Mostofa and Norick, Brandon and LeGresley, Patrick and Rajbhandari, Samyam and Casper, Jared and Liu, Zhun and Prabhumoye, Shrimai and Zerveas, George and Korthikanti, Vijay and others},
  journal={arXiv preprint arXiv:2201.11990},
  year={2022}
}

@article{adler2024nemotron,
  title={Nemotron-4 340b technical report},
  author={Adler, Bo and Agarwal, Niket and Aithal, Ashwath and Anh, Dong H and Bhattacharya, Pallab and Brundyn, Annika and Casper, Jared and Catanzaro, Bryan and Clay, Sharon and Cohen, Jonathan and others},
  journal={arXiv preprint arXiv:2406.11704},
  year={2024}
}

@article{huang2019gpipe,
  title={Gpipe: Efficient training of giant neural networks using pipeline parallelism},
  author={Huang, Yanping and Cheng, Youlong and Bapna, Ankur and Firat, Orhan and Chen, Dehao and Chen, Mia and Lee, HyoukJoong and Ngiam, Jiquan and Le, Quoc V and Wu, Yonghui and others},
  journal={Advances in neural information processing systems},
  volume={32},
  year={2019}
}

@inproceedings{narayanan2019pipedream,
  title={PipeDream: Generalized pipeline parallelism for DNN training},
  author={Narayanan, Deepak and Harlap, Aaron and Phanishayee, Amar and Seshadri, Vivek and Devanur, Nikhil R and Ganger, Gregory R and Gibbons, Phillip B and Zaharia, Matei},
  booktitle={Proceedings of the 27th ACM symposium on operating systems principles},
  pages={1--15},
  year={2019}
}

@inproceedings{narayanan2021efficient,
  title={Efficient large-scale language model training on gpu clusters using megatron-lm},
  author={Narayanan, Deepak and Shoeybi, Mohammad and Casper, Jared and LeGresley, Patrick and Patwary, Mostofa and Korthikanti, Vijay and Vainbrand, Dmitri and Kashinkunti, Prethvi and Bernauer, Julie and Catanzaro, Bryan and others},
  booktitle={Proceedings of the international conference for high performance computing, networking, storage and analysis},
  pages={1--15},
  year={2021}
}

@inproceedings{kim2023bpipe,
  title={Bpipe: Memory-balanced pipeline parallelism for training large language models},
  author={Kim, Taebum and Kim, Hyoungjoo and Yu, Gyeong-In and Chun, Byung-Gon},
  booktitle={International Conference on Machine Learning},
  pages={16639--16653},
  year={2023},
  organization={PMLR}
}

@article{peng2025dawnpiper,
  title={DawnPiper: A Memory-scablable Pipeline Parallel Training Framework},
  author={Peng, Xuan and Shi, Xuanhua and Zhang, Haolin and Zhao, Yunfei and Qian, Xuehai},
  journal={arXiv preprint arXiv:2505.05856},
  year={2025}
}

@article{arfeen2025pipefill,
  title={Pipefill: Using gpus during bubbles in pipeline-parallel llm training},
  author={Arfeen, Daiyaan and Zhang, Zhen and Fu, Xinwei and Ganger, Gregory and Wang, Yida},
  journal={Proceedings of Machine Learning and Systems},
  volume={7},
  year={2025}
}

@article{agrawal2023sarathi,
  title={Sarathi: Efficient llm inference by piggybacking decodes with chunked prefills},
  author={Agrawal, Amey and Panwar, Ashish and Mohan, Jayashree and Kwatra, Nipun and Gulavani, Bhargav S and Ramjee, Ramachandran},
  journal={arXiv preprint arXiv:2308.16369},
  year={2023}
}

@inproceedings{agrawal2024taming,
  title={Taming $\{$Throughput-Latency$\}$ tradeoff in $\{$LLM$\}$ inference with $\{$Sarathi-Serve$\}$},
  author={Agrawal, Amey and Kedia, Nitin and Panwar, Ashish and Mohan, Jayashree and Kwatra, Nipun and Gulavani, Bhargav and Tumanov, Alexey and Ramjee, Ramachandran},
  booktitle={18th USENIX Symposium on Operating Systems Design and Implementation (OSDI 24)},
  pages={117--134},
  year={2024}
}

@inproceedings{guo2025gllm,
  title={gLLM: Global Balanced Pipeline Parallelism Systems for Distributed LLMs Serving with Token Throttling},
  author={Guo, Tianyu and Zhang, Xianwei and Du, Jiangsu and Chen, Zhiguang and Xiao, Nong and Lu, Yutong},
  booktitle={Proceedings of the International Conference for High Performance Computing, Networking, Storage and Analysis},
  pages={1725--1741},
  year={2025}
}

@inproceedings{zhang2025td,
  title={TD-Pipe: Temporally-Disaggregated Pipeline Parallelism Architecture for High-Throughput LLM Inference},
  author={Zhang, Hongbin and Wei, Taosheng and Zheng, Zhenyi and Du, Jiangsu and Chen, Zhiguang and Lu, Yutong},
  booktitle={Proceedings of the 54th International Conference on Parallel Processing},
  pages={689--698},
  year={2025}
}

@article{su2025seesaw,
  title={Seesaw: High-throughput llm inference via model re-sharding},
  author={Su, Qidong and Zhao, Wei and Li, Xin and Andoorveedu, Muralidhar and Jiang, Chenhao and Zhu, Zhanda and Song, Kevin and Giannoula, Christina and Pekhimenko, Gennady},
  journal={arXiv preprint arXiv:2503.06433},
  year={2025}
}

@misc{jiang2023mistral7b,
      title={Mistral 7B}, 
      author={Albert Q. Jiang and Alexandre Sablayrolles and Arthur Mensch and Chris Bamford and Devendra Singh Chaplot and Diego de las Casas and Florian Bressand and Gianna Lengyel and Guillaume Lample and Lucile Saulnier and Lélio Renard Lavaud and Marie-Anne Lachaux and Pierre Stock and Teven Le Scao and Thibaut Lavril and Thomas Wang and Timothée Lacroix and William El Sayed},
      year={2023},
      eprint={2310.06825},
      archivePrefix={arXiv},
      primaryClass={cs.CL},
      url={https://arxiv.org/abs/2310.06825}, 
}

@misc{jiang2024mixtralexperts,
      title={Mixtral of Experts}, 
      author={Albert Q. Jiang and Alexandre Sablayrolles and Antoine Roux and Arthur Mensch and Blanche Savary and Chris Bamford and Devendra Singh Chaplot and Diego de las Casas and Emma Bou Hanna and Florian Bressand and Gianna Lengyel and Guillaume Bour and Guillaume Lample and Lélio Renard Lavaud and Lucile Saulnier and Marie-Anne Lachaux and Pierre Stock and Sandeep Subramanian and Sophia Yang and Szymon Antoniak and Teven Le Scao and Théophile Gervet and Thibaut Lavril and Thomas Wang and Timothée Lacroix and William El Sayed},
      year={2024},
      eprint={2401.04088},
      archivePrefix={arXiv},
      primaryClass={cs.LG},
      url={https://arxiv.org/abs/2401.04088}, 
}

@misc{lieber2024jambahybridtransformermambalanguage,
      title={Jamba: A Hybrid Transformer-Mamba Language Model}, 
      author={Opher Lieber and Barak Lenz and Hofit Bata and Gal Cohen and Jhonathan Osin and Itay Dalmedigos and Erez Safahi and Shaked Meirom and Yonatan Belinkov and Shai Shalev-Shwartz and Omri Abend and Raz Alon and Tomer Asida and Amir Bergman and Roman Glozman and Michael Gokhman and Avashalom Manevich and Nir Ratner and Noam Rozen and Erez Shwartz and Mor Zusman and Yoav Shoham},
      year={2024},
      eprint={2403.19887},
      archivePrefix={arXiv},
      primaryClass={cs.CL},
      url={https://arxiv.org/abs/2403.19887}, 
}

@article{blakeman2025nemotron,
  title={Nemotron-h: A family of accurate and efficient hybrid mamba-transformer models},
  author={Blakeman, Aaron and Basant, Aarti and Khattar, Abhinav and Renduchintala, Adithya and Bercovich, Akhiad and Ficek, Aleksander and Bjorlin, Alexis and Taghibakhshi, Ali and Deshmukh, Amala Sanjay and Mahabaleshwarkar, Ameya Sunil and others},
  journal={arXiv preprint arXiv:2504.03624},
  year={2025}
}

@article{lan2019albert,
  title={Albert: A lite bert for self-supervised learning of language representations},
  author={Lan, Zhenzhong and Chen, Mingda and Goodman, Sebastian and Gimpel, Kevin and Sharma, Piyush and Soricut, Radu},
  journal={arXiv preprint arXiv:1909.11942},
  year={2019}
}

@inproceedings{zuhri2025mlkv,
  title={Mlkv: Multi-layer key-value heads for memory efficient transformer decoding},
  author={Zuhri, Zayd Muhammad Kawakibi and Adilazuarda, Muhammad Farid and Purwarianti, Ayu and Aji, Alham Fikri},
  booktitle={Findings of the Association for Computational Linguistics: NAACL 2025},
  pages={5516--5525},
  year={2025}
}

@inproceedings{wu2025hshare,
  title={HShare: Fast LLM decoding by hierarchical key-value sharing},
  author={Wu, Huaijin and Li, Lianqiang and Huang, Hantao and Yi, Tu and Zhang, Jihang and Yu, Minghui and Yan, Junchi},
  booktitle={The Thirteenth International Conference on Learning Representations},
  year={2025}
}

@inproceedings{wu2025systematic,
  title={A systematic study of cross-layer kv sharing for efficient llm inference},
  author={Wu, You and Wu, Haoyi and Tu, Kewei},
  booktitle={Proceedings of the 2025 Conference of the Nations of the Americas Chapter of the Association for Computational Linguistics: Human Language Technologies (Volume 2: Short Papers)},
  pages={396--403},
  year={2025}
}

@inproceedings{kwon2023efficient,
  title={Efficient memory management for large language model serving with pagedattention},
  author={Kwon, Woosuk and Li, Zhuohan and Zhuang, Siyuan and Sheng, Ying and Zheng, Lianmin and Yu, Cody Hao and Gonzalez, Joseph and Zhang, Hao and Stoica, Ion},
  booktitle={Proceedings of the 29th symposium on operating systems principles},
  pages={611--626},
  year={2023}
}

@misc{fan2020dapplepipelineddataparallel,
      title={DAPPLE: A Pipelined Data Parallel Approach for Training Large Models}, 
      author={Shiqing Fan and Yi Rong and Chen Meng and Zongyan Cao and Siyu Wang and Zhen Zheng and Chuan Wu and Guoping Long and Jun Yang and Lixue Xia and Lansong Diao and Xiaoyong Liu and Wei Lin},
      year={2020},
      eprint={2007.01045},
      archivePrefix={arXiv},
      primaryClass={cs.DC},
      url={https://arxiv.org/abs/2007.01045}, 
}

@inproceedings{Jiang_2024dynapipe, series={EuroSys ’24},
   title={DynaPipe: Optimizing Multi-task Training through Dynamic Pipelines},
   url={http://dx.doi.org/10.1145/3627703.3629585},
   DOI={10.1145/3627703.3629585},
   booktitle={Proceedings of the Nineteenth European Conference on Computer Systems},
   publisher={ACM},
   author={Jiang, Chenyu and Jia, Zhen and Zheng, Shuai and Wang, Yida and Wu, Chuan},
   year={2024},
   month=apr, pages={542–559},
   collection={EuroSys ’24} }

@misc{li2025slimpipememorythriftyefficientpipeline,
      title={SlimPipe: Memory-Thrifty and Efficient Pipeline Parallelism for Long-Context LLM Training}, 
      author={Zhouyang Li and Yuliang Liu and Wei Zhang and Tailing Yuan and Bin Chen and Chengru Song and Di Zhang},
      year={2025},
      eprint={2504.14519},
      archivePrefix={arXiv},
      primaryClass={cs.LG},
      url={https://arxiv.org/abs/2504.14519}, 
}

@inproceedings{ao2025seq1f1b,
  title={Seq1f1b: Efficient sequence-level pipeline parallelism for large language model training},
  author={Ao, Sun and Zhao, Weilin and Han, Xu and Yang, Cheng and Zhang, Xinrong and Liu, Zhiyuan and Shi, Chuan and Sun, Maosong},
  booktitle={Proceedings of the 2025 Conference of the Nations of the Americas Chapter of the Association for Computational Linguistics: Human Language Technologies (Volume 1: Long Papers)},
  pages={8998--9008},
  year={2025}
}

@article{harlap1806pipedream,
  title={Pipedream: Fast and efficient pipeline parallel dnn training, 2018},
  author={Harlap, Aaron and Narayanan, Deepak and Phanishayee, Amar and Seshadri, Vivek and Devanur, Nikhil and Ganger, Greg and Gibbons, Phil},
  journal={URL https://arxiv. org/abs},
  year={1806}
}

@inproceedings{dialameh2025echo,
  title={ECHO-LLaMA: Efficient Caching for High-Performance LLaMA Training},
  author={Dialameh, Maryam and Karim, Rezaul and Rajabzadeh, Hossein and Awad, Omar Mohamed and Chen, Boxing and Kwon, Hyock Ju and Ahmed, Walid and Liu, Yang},
  booktitle={Proceedings of the 2025 Conference on Empirical Methods in Natural Language Processing: Industry Track},
  pages={2252--2269},
  year={2025}
}

@article{rajabzadeh2026efficient,
  title={Efficient Learning for Large Language Models},
  author={Rajabzadeh, Hossein},
  year={2026},
  publisher={University of Waterloo}
}

\newpage
\appendix
\onecolumn

\section{Additional Analysis and Validation}
\label{app:additional_analysis}

This appendix provides detailed analyses and additional experiments
that complement the main evaluation. We organize the material around the
questions that arise naturally from KV-Pipe's design:
(i) how different KV-sharing placements reshape stage-level compute,
(ii) whether the selected configurations preserve model quality,
(iii) how sensitive the method is to placement, the FIR diagnostic, and the
stopping tolerance,
(iv) whether KV-Pipe remains useful on top of stronger pipeline schedules and
prior partitioning methods, and
(v) whether the inference-side benefit persists under grouped-query attention
(GQA). We conclude with practical guidance and the scope of the claims supported
by the current experiments.

\subsection{Motivating Example: Architectural Stage Imbalance} \label{app:baseline_split} Pipeline stages can remain imbalanced even when the Transformer layers themselves are evenly partitioned. This occurs because architectural components such as the embedding layer and LM head are not uniformly distributed across stages. As an illustrative example, consider LLaMA2-7B with $L=32$ attention layers and $P=8$ pipeline stages. Table~\ref{table:baseline} shows a baseline partition with four attention layers per stage. Despite the uniform layer assignment, the last stage has substantially higher FLOPs and forward time because it additionally contains the LM head. \begin{table*}[t] \centering \small \setlength{\tabcolsep}{3pt} \begin{tabular}{c|cccccccc} \hline & 0 & 1 & 2 & 3 & 4 & 5 & 6 & 7 \\ \hline Emb & 1 & 0 & 0 & 0 & 0 & 0 & 0 & 0 \\ Full attn & 4 & 4 & 4 & 4 & 4 & 4 & 4 & 4 \\ KV sharing & 0 & 0 & 0 & 0 & 0 & 0 & 0 & 0 \\ LM head & 0 & 0 & 0 & 0 & 0 & 0 & 0 & 1 \\ \hline FLOPs & 1.55E+13 & 1.55E+13 & 1.55E+13 & 1.55E+13 & 1.55E+13 & 1.55E+13 & 1.55E+13 & 1.763E+13 \\ FWD time & 59712.13 & 59237.7 & 59237.7 & 59237.7 & 59237.7 & 59237.7 & 59237.7 & 66796.0045 \\ Memory & 16.8425 & 15.5625 & 15.5625 & 15.5625 & 15.5625 & 15.5625 & 15.5625 & 16.8425 \\ \hline \end{tabular} \caption{\textbf{Baseline configuration} for LLaMA2-7B with PP=8, sequence length 8K, and MBS=1. Although the attention layers are evenly partitioned, the final stage becomes the bottleneck because it also contains the LM head.} \label{table:baseline} 
\end{table*} 

\subsection{Full-Attention and KV-Sharing Layers} \label{app:kvsharing_details} For hidden states $\mathbf{X}_{\ell}\in\mathbb{R}^{S\times d}$, a standard attention layer computes \begin{equation} \mathbf{Q}_{\ell}=\mathbf{X}_{\ell}\mathbf{W}_{\ell}^{Q},\qquad \mathbf{K}_{\ell}=\mathbf{X}_{\ell}\mathbf{W}_{\ell}^{K},\qquad \mathbf{V}_{\ell}=\mathbf{X}_{\ell}\mathbf{W}_{\ell}^{V}, \end{equation} followed by \begin{equation} \mathbf{O}_{\ell} = \mathrm{softmax}\!\left( \frac{\mathbf{Q}_{\ell}\mathbf{K}_{\ell}^{\top}}{\sqrt{d_h}} +\mathbf{M} \right)\mathbf{V}_{\ell}, \label{eq:attn_full_appendix} \end{equation} where $\mathbf{M}$ is the causal mask and $d_h$ is the per-head dimension. Cross-layer KV sharing replaces the layer-specific key/value tensors with those of an earlier layer: \begin{equation} \mathbf{K}_{\ell}\leftarrow\mathbf{K}_{\tau(\ell)},\qquad \mathbf{V}_{\ell}\leftarrow\mathbf{V}_{\tau(\ell)}. \end{equation} The query projection and attention operation remain unchanged. Therefore, each converted layer reduces both projection FLOPs and the amount of additional KV state produced by that layer. KV-Pipe exploits the former to redistribute pipeline computation and benefits from the latter during long-context inference. \subsection{KV-Sharing Placement Strategies} \label{app:strategies} 
\begin{figure}[t] \centering \includegraphics[width=0.6\linewidth]{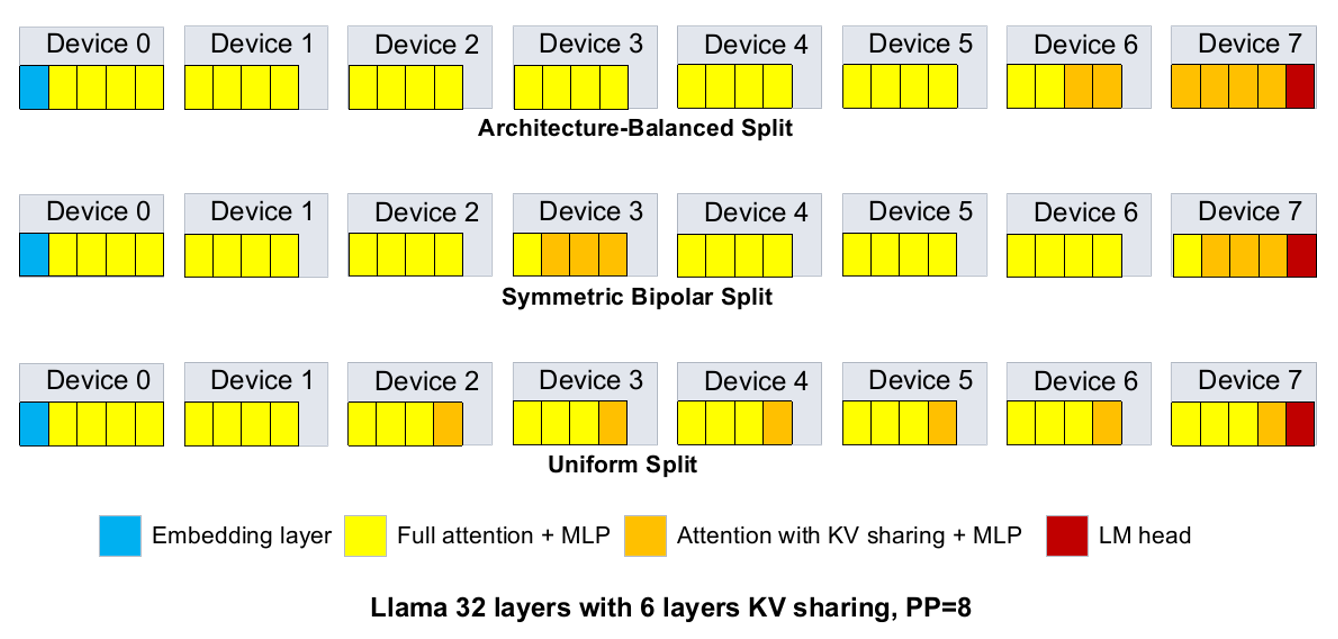} \caption{Illustration of the evaluated KV-sharing placement strategies for a 32-layer LLaMA model under pipeline parallelism. Uniform distributes sharing broadly, Symmetric Bipolar uses two allocation anchors, and Architecture-Balanced concentrates sharing on bottleneck stages.} \label{fig:kv_strategies} \end{figure}

We evaluate three representative placement policies, illustrated in Figure~\ref{fig:kv_strategies}. They differ only in \emph{where} the available KV-sharing layers are placed, allowing us to isolate the importance of stage-aware allocation. \paragraph{Uniform.} KV-sharing layers are distributed approximately evenly across pipeline stages. This reduces total computation but does not explicitly target the bottleneck. \paragraph{Symmetric Bipolar.} KV-sharing layers are allocated from two anchors---the tail stage and a middle stage---and expanded toward earlier stages. This policy distributes computation reduction across two regions of the pipeline and is useful when the imbalance is not concentrated entirely at the tail. \paragraph{Architecture-Balanced.} KV-sharing layers are preferentially assigned to the current bottleneck stage. The policy initially targets the final stage and subsequently follows the maximum-FLOP stage as the bottleneck shifts. This is the placement policy used by KV-Pipe.  \subsection{Architecture-Balanced KV-Pipe Algorithm} \label{app:kvpipe_algorithm} Algorithm~\ref{alg:kvpipe} gives the complete architecture-balanced procedure. KV-Pipe starts from the final pipeline stage and converts its deepest remaining full-attention layer to KV sharing. This tail-first policy is motivated by the common case in which the LM head makes the final stage more expensive than the others. After each conversion, the algorithm decreases the corresponding stage FLOPs by \begin{equation} \Delta F_{\ell^\star} = F_{\ell^\star}(0)-F_{\ell^\star}(1), \end{equation} and recomputes FIR. Once the final stage falls within the specified tolerance of the average stage cost, KV-Pipe selects the current bottleneck $i^\star=\arg\max_i F_i$. Thus, sharing follows the bottleneck if it migrates to another stage. The algorithm terminates once the pipeline is sufficiently balanced ($\mathrm{FIR}\le 1+\epsilon$), the sharing budget $m$ is exhausted, or no eligible layer remains. For at most $m$ conversions, bottleneck identification requires only a scan over the $P$ stages, giving a negligible $O(mP)$ offline cost for practical PP configurations. 

\begin{algorithm*}[t] 
\caption{KV-Pipe: Architecture-Balanced KV Sharing} \label{alg:kvpipe} \small \begin{algorithmic}[1] \Require Pipeline partition $\{\mathcal{S}_i\}_{i=1}^{P}$; full-attention and KV-sharing FLOPs $\{F_\ell(0),F_\ell(1)\}_{\ell=1}^{L}$; stage-specific costs $\{C_i\}_{i=1}^{P}$; tolerance $\epsilon$; maximum number of conversions $m$. \Ensure KV-sharing mask $\mathbf{z}$ and resulting FIR. \State $\mathbf{z}\leftarrow\mathbf{0}$ \State $F_i\leftarrow \sum_{\ell\in\mathcal{S}_i}F_\ell(0)+C_i,\quad \forall i$ \State Compute $\mathrm{FIR}$ using Eq.~\eqref{eq:flop_imbalance_main} \State $t\leftarrow0$, $i^\star\leftarrow P$ \Comment{start from the tail stage} \While{$\mathrm{FIR}>1+\epsilon$ \textbf{and} $t<m$} \If{$F_{i^\star}\le(1+\epsilon)F_{\mathrm{avg}}$} \State $i^\star\leftarrow\arg\max_j F_j$ \Comment{follow the current bottleneck} \EndIf \State $\ell^\star\leftarrow \max\{\ell\in\mathcal{S}_{i^\star}:z_\ell=0\}$ \Comment{last convertible layer} \If{$\ell^\star$ is undefined} \State \textbf{break} \EndIf \State $z_{\ell^\star}\leftarrow1$ \State $\Delta F_{\ell^\star}\leftarrow F_{\ell^\star}(0)-F_{\ell^\star}(1)$ \State $F_{i^\star}\leftarrow F_{i^\star}-\Delta F_{\ell^\star}$ \State $t\leftarrow t+1$ \State Recompute $F_{\mathrm{avg}}$ and $\mathrm{FIR}$ \EndWhile \State \Return $\mathbf{z}$, $\{F_i\}_{i=1}^{P}$, $\mathrm{FIR}$ 
\end{algorithmic} 
\end{algorithm*} 

\subsection{Illustrative FIR Comparison} \label{app:fir_case_study} Table~\ref{table:Imbalancerratio} illustrates the effect of placement using LLaMA2-7B with four KV-sharing layers, PP=8, sequence length 8K, and MBS=1. Simply introducing KV sharing improves the imbalance under all three policies, but its placement has a substantial effect. Uniform sharing decreases FIR only from $1.1195$ to $1.1038$, while Symmetric Bipolar reaches $1.0683$. Architecture-Balanced placement reduces FIR to $1.0004$, yielding an almost perfect FLOP balance. \begin{table}[t] \centering \small \begin{tabular}{l|c} \hline \textbf{PP split strategy} & \textbf{FIR} \\ \hline Baseline & 1.1195 \\ Uniform & 1.1038 \\ Symmetric Bipolar & 1.0683 \\ Architecture-Balanced & \textbf{1.0004} \\ \hline \end{tabular} \caption{FLOPs Imbalance Ratio for LLaMA2-7B with four KV-sharing layers, PP=8, sequence length 8K, and MBS=1. Values closer to $1$ indicate better pipeline balance.} \label{table:Imbalancerratio} \end{table}

\subsection{Stage-Level Effect of KV-Sharing Placement}
\label{app:stage_level_splits}

We first make the mechanism of KV-Pipe explicit at the stage level. The main
paper shows that, for LLaMA2-7B with PP=8 and $S=8$K, the last stage is
initially the bottleneck because it contains both Transformer blocks and the LM
head. Tables~\ref{table:uniform}--\ref{table:architecture-balanced} show how the
same SKV budget is distributed by the three placement strategies considered in
the paper.

The important distinction is \emph{where} the compute reduction is applied.
Uniform placement spreads the SKV budget across several stages; Symmetric
Bipolar concentrates it at two locations; Architecture-Balanced concentrates
the budget on the current critical stage. As a result, Architecture-Balanced
brings the tail-stage FLOPs close to the other stages and nearly equalizes the
stage-time profile. This stage-level view explains why reducing total FLOPs
alone is not sufficient: reductions that do not lie on the critical path can
leave the pipeline bottleneck largely unchanged.

\begin{table*}[t]
\centering
\small
\setlength{\tabcolsep}{3pt}
\begin{tabular}{c|cccccccc}
\toprule
 & 0 & 1 & 2 & 3 & 4 & 5 & 6 & 7 \\
\midrule
Emb & 1 & 0 & 0 & 0 & 0 & 0 & 0 & 0 \\
Full attn & 4 & 4 & 4 & 4 & 3 & 3 & 3 & 3 \\
KV sharing & 0 & 0 & 0 & 0 & 1 & 1 & 1 & 1 \\
LM head & 0 & 0 & 0 & 0 & 0 & 0 & 0 & 1 \\
\midrule
FLOPs & 1.55E+13 & 1.55E+13 & 1.55E+13 & 1.55E+13 &
1.49E+13 & 1.49E+13 & 1.49E+13 & 1.708E+13 \\
FWD time & 59712.13 & 59237.7 & 59237.7 & 59237.7 &
57301.7 & 57301.7 & 57301.7 & 64860.0028 \\
Memory & 16.8425 & 15.5625 & 15.5625 & 15.5625 &
15.5597 & 15.5597 & 15.5597 & 16.8395703 \\
\bottomrule
\end{tabular}
\caption{\textbf{Uniform placement} for LLaMA2-7B, PP=8, $S=8$K, MBS=1.
The SKV budget is spread across stages 4--7; stage 7 remains the bottleneck.}
\label{table:uniform}
\end{table*}

\begin{table*}[t]
\centering
\small
\setlength{\tabcolsep}{3pt}
\begin{tabular}{c|cccccccc}
\toprule
 & 0 & 1 & 2 & 3 & 4 & 5 & 6 & 7 \\
\midrule
Emb & 1 & 0 & 0 & 0 & 0 & 0 & 0 & 0 \\
Full attn & 4 & 4 & 4 & 2 & 4 & 4 & 4 & 2 \\
KV sharing & 0 & 0 & 0 & 2 & 0 & 0 & 0 & 2 \\
LM head & 0 & 0 & 0 & 0 & 0 & 0 & 0 & 1 \\
\midrule
FLOPs & 1.55E+13 & 1.55E+13 & 1.55E+13 & 1.44E+13 &
1.55E+13 & 1.55E+13 & 1.55E+13 & 1.653E+13 \\
FWD time & 59712.13 & 59237.7 & 59237.7 & 55365.7 &
59237.7 & 59237.7 & 59237.7 & 62924.0012 \\
Memory & 16.8425 & 15.5625 & 15.5625 & 15.55664 &
15.5625 & 15.5625 & 15.5625 & 16.8366406 \\
\bottomrule
\end{tabular}
\caption{\textbf{Symmetric-Bipolar placement} for LLaMA2-7B, PP=8,
$S=8$K, MBS=1. The SKV budget is split between stages 3 and 7; stage 7
still has the largest FLOPs and forward time.}
\label{table:symmetric-bipolar}
\end{table*}

\begin{table*}[t]
\centering
\small
\setlength{\tabcolsep}{3pt}
\begin{tabular}{c|cccccccc}
\toprule
 & 0 & 1 & 2 & 3 & 4 & 5 & 6 & 7 \\
\midrule
Emb & 1 & 0 & 0 & 0 & 0 & 0 & 0 & 0 \\
Full attn & 4 & 4 & 4 & 4 & 4 & 4 & 4 & 0 \\
KV sharing & 0 & 0 & 0 & 0 & 0 & 0 & 0 & 4 \\
LM head & 0 & 0 & 0 & 0 & 0 & 0 & 0 & 1 \\
\midrule
FLOPs & 1.55E+13 & 1.55E+13 & 1.55E+13 & 1.55E+13 &
1.55E+13 & 1.55E+13 & 1.55E+13 & 1.543E+13 \\
FWD time & 59712.13 & 59237.7 & 59237.7 & 59237.7 &
59237.7 & 59237.7 & 59237.7 & 59051.9978 \\
Memory & 16.8425 & 15.5625 & 15.5625 & 15.5625 &
15.5625 & 15.5625 & 15.5625 & 16.8307812 \\
\bottomrule
\end{tabular}
\caption{\textbf{Architecture-Balanced placement} for LLaMA2-7B, PP=8,
$S=8$K, MBS=1. All four SKV conversions are applied to the original tail
bottleneck. The resulting stage FLOPs are nearly equal, after which stage 0
becomes marginally critical in measured forward time.}
\label{table:architecture-balanced}
\end{table*}

\subsection{Quality--Efficiency Trade-off and Final Layer Assignments}
\label{app:quality_efficiency}

Cross-layer KV sharing permanently changes the model architecture, so system
efficiency must be interpreted together with model quality. We therefore
evaluate validation perplexity and downstream accuracy for the exact
configurations selected by KV-Pipe and compare them with full attention and a
fixed Echo-style 25\% tail-sharing baseline.

Table~\ref{tab:app_quality_efficiency} also reports the final SKV assignments.
For LLaMA2-7B, KV-Pipe selects layers 25--32 at PP=2, 27--32 at PP=4, and
29--32 at PP=8. Hence, in every quality-tested configuration, the earliest
converted layer remains in the second half of the 32-layer network. We stress
that this is an empirical observation for the evaluated configurations, not a
theoretical guarantee that the bottleneck can never move to an earlier stage in
an arbitrary model or PP layout.

\begin{table*}[t]
\centering
\small
\setlength{\tabcolsep}{3.5pt}
\renewcommand{\arraystretch}{1.08}
\caption{
\textbf{Quality--efficiency Pareto for LLaMA2-7B.}
Efficiency changes are reported relative to the corresponding full-attention
1F1B baseline. SKV denotes the number of attention layers converted to
shared-KV layers; layer indices are 1-indexed. Downstream Avg.\ is the average
over MMLU, HellaSwag, ARC-Challenge, and TruthfulQA.
}
\label{tab:app_quality_efficiency}
\label{tab:review_response_quality_efficiency}
\resizebox{\textwidth}{!}{
\begin{tabular}{llccccccccc}
\toprule
\textbf{Setting} &
\textbf{Method} &
\textbf{PP} &
\textbf{Seq.} &
\textbf{SKV} &
\textbf{Converted layers} &
\textbf{Earliest} &
\textbf{Late-half?} &
\textbf{Val. PPL $\downarrow$} &
\textbf{Downstream Avg. $\uparrow$} &
\textbf{MFU / Iter. time change} \\
\midrule

$S{=}4$K
& Full-attn baseline
& 2 & 4K & 0 & -- & -- & --
& 5.54 & 52.2 & 0.00\% / 0.00\% \\
& Echo-style 25\% sharing
& 2 & 4K & 8 & 25--32 & 25 & Yes
& 5.51 & 51.9 & +2.85\% / -4.35\% \\
& KV-Pipe, optimal
& 2 & 4K & 8 & 25--32 & 25 & Yes
& 5.51 & 51.9 & \textbf{+3.10\% / -4.90\%} \\

\midrule

$S{=}8$K
& Full-attn baseline
& 4 & 8K & 0 & -- & -- & --
& 5.67 & 52.5 & 0.00\% / 0.00\% \\
& Echo-style 25\% sharing
& 4 & 8K & 8 & 25--32 & 25 & Yes
& 5.50 & 52.0 & +3.70\% / -6.35\% \\
& KV-Pipe, optimal
& 4 & 8K & 6 & 27--32 & 27 & Yes
& 5.50 & 52.0 & \textbf{+4.32\% / -7.56\%} \\

\midrule

$S{=}8$K
& Full-attn baseline
& 8 & 8K & 0 & -- & -- & --
& 5.47 & 52.2 & 0.00\% / 0.00\% \\
& Echo-style 25\% sharing
& 8 & 8K & 8 & 25--32 & 25 & Yes
& 5.53 & 52.0 & +6.80\% / -8.10\% \\
& KV-Pipe, optimal
& 8 & 8K & 4 & 29--32 & 29 & Yes
& 5.49 & 52.2 & \textbf{+9.17\% / -9.80\%} \\

\bottomrule
\end{tabular}
}
\end{table*}

\paragraph{Fixed Echo-style sharing versus adaptive budget selection.}
The Echo-style baseline always converts the last 25\% of LLaMA2-7B, i.e.,
layers 25--32. KV-Pipe instead chooses how much sharing is required from the
pipeline imbalance. At PP=8 and $S=8$K, KV-Pipe uses four SKV layers rather
than eight, obtains validation PPL 5.49 versus 5.53 and downstream average
52.2 versus 52.0, and reduces iteration time by 9.80\% rather than 8.10\%.
The practical advantage in this setting is therefore a better
\emph{quality--sharing-budget--efficiency} operating point, rather than a claim
that more KV sharing is universally worse.

At PP=2, the fixed 25\% rule and KV-Pipe select the same layers (25--32), and
their system results are correspondingly close. The difference becomes more
meaningful at higher PP, where KV-Pipe stops earlier once the stage imbalance
has been sufficiently reduced.

\paragraph{Matched-budget interpretation.}
For the evaluated LLaMA2 settings, Architecture-Balanced placement remains
tail-localized. If a fixed tail-sharing heuristic is given the \emph{same}
budget selected by KV-Pipe, it therefore selects the same layer sets:
25--32 for SKV=8, 27--32 for SKV=6, and 29--32 for SKV=4.
These experiments support a more precise interpretation of KV-Pipe: in these
LLaMA2 cases, its main benefit is to determine the SKV budget from the measured
pipeline imbalance instead of fixing a 25\% or 50\% sharing ratio a priori.
The bottleneck-retargeting mechanism is designed for the more general case in
which the critical stage moves away from the tail.

\subsection{Optional Safe-Layer Guardrail}
\label{app:safe_guard}

Although the tail-first policy stays in late layers in all quality-tested
configurations above, KV-Pipe optimizes a systems objective and does not imply
that early-layer conversion is universally quality-neutral. To make the
quality constraint explicit, KV-Pipe can be run with a user-controlled
admissible layer set.

Let $\alpha\in[0,1)$ denote the minimum normalized depth at which KV sharing is
allowed. We define
\begin{equation}
\mathcal{C}_{\alpha}
=
\left\{
\ell \in \{1,\ldots,L\}
:
\ell \ge \left\lceil \alpha L \right\rceil
\right\}.
\end{equation}
The layer-selection step is then restricted to
\begin{equation}
\ell^\star
=
\max
\left\{
\ell \in
\mathcal{S}_{i^\star}\cap\mathcal{C}_{\alpha}
:
z_\ell=0
\right\}.
\end{equation}

For example, $\alpha=0.5$ restricts conversions to the second half of the
network. If no admissible layer remains before the target FIR tolerance is
reached, KV-Pipe terminates and reports the residual imbalance rather than
converting an earlier layer. This guard does not alter any result in
Table~\ref{tab:app_quality_efficiency}, because all selected layers already lie
in the second half of LLaMA2-7B.

\subsection{Sensitivity to Placement, FIR, and the Stopping Tolerance}
\label{app:sensitivity}

We next isolate three design choices: the placement policy, the stopping
tolerance $\epsilon$, and the imbalance signal. All rows in
Table~\ref{tab:app_sensitivity} use LLaMA2-7B with PP=8 and $S=8$K under the
same 1F1B setting as the NPU experiments.

\begin{table*}[t]
\centering
\small
\setlength{\tabcolsep}{5pt}
\caption{
\textbf{Sensitivity and diagnostic ablations.}
MFU and iteration-time changes are relative to the full-attention 1F1B
baseline at PP=8 and $S=8$K.
}
\label{tab:app_sensitivity}
\label{tab:sensitivity_diagnostic}
\resizebox{\textwidth}{!}{
\begin{tabular}{llclccc}
\toprule
\textbf{Ablation} &
\textbf{Setting} &
\textbf{SKV} &
\textbf{Converted layers} &
\textbf{Val. PPL $\downarrow$} &
\textbf{Downstream Avg. $\uparrow$} &
\textbf{MFU / Iter. time change} \\
\midrule

Uniform placement
& PP=8, $S$=8K
& 4
& 17, 22, 27, 32
& 5.50
& 52.1
& +0.9\% / -2.4\%  \\

Symmetric bipolar
& PP=8, $S$=8K
& 4
& 17, 18, 31, 32
& 5.50
& 52.1
& +3.2\% / -4.7\% \\

Architecture-Balanced
& PP=8, $S$=8K
& 4
& 29--32
& 5.49
& 52.2
& \textbf{+9.17\% / -9.80\%} \\

\midrule

$\epsilon=0.02$
& PP=8, $S$=8K
& 5
& 28--32
& 5.50
& 52.1
& +9.05\% / -9.65\% \\

$\epsilon=0.05$
& PP=8, $S$=8K
& 4
& 29--32
& 5.49
& 52.2
& \textbf{+9.17\% / -9.80\%} \\

$\epsilon=0.10$
& PP=8, $S$=8K
& 3
& 30--32
& 5.48
& 52.2
& +8.60\% / -9.20\% \\

\midrule

FIR-guided selection
& PP=8, $S$=8K
& 4
& 29--32
& 5.49
& 52.2
& \textbf{+9.17\% / -9.80\%} \\

Max-stage-time guided
& PP=8, $S$=8K
& 4
& 28--31
& 5.49
& 52.2
& +8.90\% / -9.45\% \\

\bottomrule
\end{tabular}
}
\end{table*}

\paragraph{Placement policy.}
At the same SKV=4 budget, Architecture-Balanced placement gives the largest
efficiency improvement among the three representative strategies. Uniform and
Symmetric Bipolar placement spend part of the SKV budget on stages that are not
always on the critical path, whereas Architecture-Balanced concentrates the
reduction on the bottleneck stage.

\paragraph{Tolerance sensitivity.}
The method is relatively stable over the tested range. Changing $\epsilon$
from 0.02 to 0.10 changes the selected budget from five to three SKV layers,
while the MFU improvement remains between 8.60\% and 9.17\% and validation PPL
remains between 5.48 and 5.50. Thus, the observed optimum does not require
fine-grained tuning of $\epsilon$ for this workload.

\paragraph{FIR versus measured stage-time guidance.}
FIR is intended as a lightweight analytic imbalance proxy, not as a claim of
universal superiority over timing-based diagnostics. In this ablation,
FIR-guided selection obtains +9.17\% MFU and -9.80\% iteration time, compared
with +8.90\% and -9.45\% for max-stage-time guidance. This supports FIR as a
useful low-cost offline signal for the evaluated setup, while direct timing may
remain preferable when accurate stage profiles are already available. The
current experiment should therefore be read as a diagnostic ablation rather
than a universal predictive-power guarantee.

\subsection{Composability with a Stronger Pipeline Schedule}
\label{app:composability}

KV-Pipe changes the per-stage workload and does not require a particular
pipeline scheduler. To test whether its benefit remains after schedule-level
bubble reduction, we combine KV-Pipe with Seq1F1B.

\begin{table}[t]
\centering
\small
\caption{
\textbf{Composability with Seq1F1B.}
KV-Pipe provides an additional gain after applying a stronger pipeline
schedule.
}
\label{tab:app_seq1f1b}
\begin{tabular}{llcc}
\toprule
\textbf{Model / Setting} &
\textbf{Method} &
\textbf{MFU} &
\textbf{Iter. time} \\
\midrule

LLaMA2-7B, PP=8, 8K
& 1F1B
& 39.0\%
& 305s \\
& Seq1F1B
& 47.0\%
& 265s \\
& Seq1F1B + KV-Pipe
& \textbf{50.0\%}
& \textbf{248s} \\

\midrule

LLaMA2-13B, PP=8, 16K
& 1F1B
& 39.5\%
& 960s \\
& Seq1F1B
& 48.5\%
& 820s \\
& Seq1F1B + KV-Pipe
& \textbf{52.0\%}
& \textbf{750s} \\

\bottomrule
\end{tabular}
\end{table}

Relative to Seq1F1B alone, adding KV-Pipe increases MFU by approximately
6.4\% on LLaMA2-7B and 7.2\% on LLaMA2-13B, while reducing iteration time by
approximately 6.4\% and 8.5\%, respectively. This provides direct evidence for
the composability claim: schedule-level bubble mitigation and stage-workload
balancing address different sources of inefficiency.

\paragraph{Hardware coverage.}
The Ascend 910B experiments use MindSpeed-LM's standard 1F1B schedule.
Seq1F1B requires schedule-level runtime support that is available in our
CUDA/GPU stack but is not available in our current Ascend 910B implementation.
We therefore evaluate Seq1F1B composability on GPUs rather than implying
identical scheduler coverage across software stacks. This mismatch is an
implementation limitation of the current evaluation, not a requirement of
KV-Pipe itself.

\subsection{Comparison with Prior Pipeline-Parallel Methods}
\label{app:partition_comparison}

Prior PP systems attack inefficiency through several different levers:
scheduling, partition search, dynamic micro-batching, memory balancing, and
bubble filling. KV-Pipe is different in that it changes the effective cost of
selected attention layers. We report both a quantitative comparison under the
vPipe/DawnPiper benchmark setting and a high-level taxonomy of representative
systems.

\subsubsection{Quantitative comparison with vPipe and DawnPiper}
\begin{table*}[t]
\centering
\small
\setlength{\tabcolsep}{3pt}

\caption{
\textbf{Training-time comparison with prior pipeline partitioning methods.}
Experiments use the DawnPiper/vPipe evaluation setting with 8 NVIDIA A100
40GB GPUs and 8 pipeline stages. Throughput is averaged over the feasible
batch-size sweep. Speedup is normalized to vPipe-AS for each model.
Stage imbalance is measured as
$\max_i t_i / \mathrm{avg}_i t_i$.
}
\label{tab:kvpipe_vs_training_partitioners}

\begin{tabularx}{\textwidth}{
@{}
l
l
>{\raggedright\arraybackslash}X
c
c
c
c
@{}
}
\toprule

\textbf{Model} &
\textbf{Method} &
\textbf{Pipeline mode} &
\shortstack{\textbf{Max}\\\textbf{batch}} &
\shortstack{\textbf{Avg. speed}\\\textbf{(samples/s)}} &
\shortstack{\textbf{Speedup}\\\textbf{vs. vPipe-AS}} &
\shortstack{\textbf{Stage}\\\textbf{imbalance}} \\

\midrule

GPT-2 770M
& GPipe
& Sync.
& 7
& 16.0
& 0.56$\times$
& 1.06 \\

& vPipe-S
& Sync.
& 7
& 16.2
& 0.56$\times$
& 1.05 \\

& vPipe-AS
& Async.
& 16
& 28.7
& 1.00$\times$
& 1.05 \\

& DawnPiper-AS
& Async.
& 20
& 33.0
& 1.15$\times$
& 1.03 \\

& KV-Pipe
& Async. + SKV rebalance
& 20
& \textbf{34.5}
& \textbf{1.20$\times$}
& \textbf{1.02} \\

\midrule

T5 780M
& GPipe
& Sync.
& 80
& 56.0
& 0.48$\times$
& 1.42 \\

& vPipe-S
& Sync.
& 80
& 62.0
& 0.53$\times$
& 1.31 \\

& vPipe-AS
& Async.
& 180
& 116.0
& 1.00$\times$
& 1.25 \\

& DawnPiper-AS
& Async.
& 220
& 155.0
& 1.34$\times$
& 1.12 \\

& KV-Pipe
& Async. + SKV rebalance
& 220
& \textbf{163.0}
& \textbf{1.41$\times$}
& \textbf{1.07} \\

\bottomrule
\end{tabularx}

\end{table*}
Compared directly with DawnPiper-AS, KV-Pipe improves average throughput by
approximately 4.5\% on GPT-2 and 5.2\% on T5, while reducing the measured
stage-imbalance ratio from 1.03 to 1.02 and from 1.12 to 1.07, respectively.
Relative to vPipe-AS, the corresponding speedups are 1.20$\times$ and
1.41$\times$. These results should be interpreted as evidence that changing the
per-layer cost profile can provide additional headroom beyond partition search,
not as a claim that KV-Pipe subsumes general PP partitioning methods.

\subsubsection{Relationship to other PP optimization levers}

Table~\ref{tab:prior_art_comparison} summarizes the role of representative PP
methods at a high level. This table is intended as a taxonomy rather than a
performance ranking. Methods that primarily change the schedule or search for a
partition can in principle be combined with KV-Pipe because KV-Pipe operates on
the model's per-layer cost profile.

\begin{table*}[t]
\centering
\small
\caption{\textbf{High-level positioning of KV-Pipe relative to representative PP optimizations.}
The table summarizes the primary optimization lever rather than claiming that
the categories are mutually exclusive.}
\label{tab:prior_art_comparison}
\begin{tabularx}{\textwidth}{l X X}
\toprule
\textbf{Method} & \textbf{Primary optimization lever} & \textbf{Relationship to KV-Pipe} \\
\midrule
\textbf{KV-Pipe} &
Stage-aware cross-layer KV sharing that changes the effective per-layer and
per-stage compute profile. &
Directly targets stage skew; can be layered with scheduling or partition-search
methods. \\

GPipe \cite{huang2019gpipe} &
Micro-batched synchronous pipeline scheduling. &
Primarily changes execution schedule; KV-Pipe can change the stage costs seen
by the schedule. \\

PipeDream \cite{narayanan2019pipedream} &
Asynchronous pipeline execution and forward/backward overlap. &
Primarily changes execution overlap; stage-cost balancing is orthogonal. \\

DAPPLE \cite{fan2020dapplepipelineddataparallel} &
Joint planning of data and pipeline parallel execution. &
Planner/search can operate on a model whose layer costs have already been
modified by KV sharing. \\

DynaPipe \cite{Jiang_2024dynapipe} &
Dynamic micro-batching and runtime adaptation. &
Runtime adaptation and offline stage-cost reshaping address different sources
of imbalance. \\

BPipe \cite{kim2023bpipe} &
Memory-oriented pipeline balancing and activation movement. &
Targets memory imbalance; KV-Pipe primarily targets compute skew while also
reducing KV-related memory. \\

SlimPipe \cite{li2025slimpipememorythriftyefficientpipeline} &
Fine-grained sequence slicing for pipeline execution. &
Sequence-level scheduling can be combined with stage-level compute reshaping. \\

PipeFill \cite{arfeen2025pipefill} &
Filling pipeline bubbles with additional work/jobs. &
Uses otherwise idle slots; KV-Pipe instead attempts to reduce the underlying
stage skew that creates idle time. \\

vPipe &
Pipeline partitioning with memory/compute considerations. &
Partition search changes where work is placed; KV-Pipe changes the cost of
selected attention layers. \\

DawnPiper \cite{peng2025dawnpiper} &
Cost-model-driven pipeline partition search. &
Complementary in objective: DawnPiper changes placement, while KV-Pipe can
reshape the cost profile supplied to a partitioner. \\

Unity / Galvatron / FlexPipe / Varuna &
Automatic or hybrid search over pipeline/parallel execution strategies. &
Potentially composable with KV-Pipe because their search space can operate on
the modified stage-cost profile. \\

\bottomrule
\end{tabularx}
\end{table*}

\subsection{Attention FLOPs under MHA and GQA}
\label{app:flops_per_token_attn}

The original inference case study uses multi-head attention (MHA), whereas many
modern LLMs use grouped-query attention (GQA). We therefore derive the
attention FLOPs for both cases and quantify what cross-layer KV sharing can
still remove under GQA. We follow the standard matrix-multiplication FLOP rule
and count a fused multiply-add (MAC) as two FLOPs.

\subsubsection{Matrix-multiplication FLOPs}

For matrices $\mathbf{A}\in\mathbb{R}^{m\times k}$ and
$\mathbf{B}\in\mathbb{R}^{k\times n}$,
\begin{equation}
\mathrm{FLOPs}(\mathbf{A}\mathbf{B}) = 2mkn.
\label{eq:mm_flops}
\end{equation}

\subsubsection{Notation}

Let sequence length be $s$, batch size be $b$, hidden size be $h$, number of
query heads be $a$, and per-head dimension be $d=h/a$. The input activations
to an attention layer are
$\mathbf{X}\in\mathbb{R}^{(bs)\times h}$.

For GQA, let $g$ be the number of KV heads and define
\begin{equation}
r \triangleq \frac{a}{g} \ge 1,
\label{eq:gqa_r}
\end{equation}
so that the total KV projection dimension is $g d=h/r$.

\subsubsection{Q/K/V projections}

\paragraph{MHA.}
MHA uses full-size projections for all three tensors:
\begin{equation}
\mathbf{Q}=\mathbf{X}\mathbf{W}^Q,\quad
\mathbf{K}=\mathbf{X}\mathbf{W}^K,\quad
\mathbf{V}=\mathbf{X}\mathbf{W}^V,
\end{equation}
with $\mathbf{W}^{(\cdot)}\in\mathbb{R}^{h\times h}$. Hence
\begin{equation}
\mathrm{FLOPs}_{\mathrm{QKV}}^{\mathrm{MHA}}
=
3\cdot 2(bs)h^2
=
6bsh^2.
\label{eq:qkv_mha}
\end{equation}

\paragraph{GQA.}
GQA keeps the query projection full-size but reduces the K/V dimension to
$h/r$:
\begin{equation}
\mathrm{FLOPs}_{\mathrm{QKV}}^{\mathrm{GQA}}
=
2(bs)h^2
+
2\cdot 2(bs)h\left(\frac{h}{r}\right)
=
2bsh^2\left(1+\frac{2}{r}\right).
\label{eq:qkv_gqa}
\end{equation}

\subsubsection{Attention score and weighted sum}

For each query head,
$\mathbf{Q}\mathbf{K}^{\top}$ costs $2s^2d$ FLOPs. Summed over the batch and
all query heads,
\begin{equation}
\mathrm{FLOPs}_{\mathrm{score}}
=
2bas^2d
=
2bs^2h.
\label{eq:qk}
\end{equation}
The weighted sum $\mathrm{Attn}\cdot\mathbf{V}$ has the same cost:
\begin{equation}
\mathrm{FLOPs}_{\mathrm{ctx}}
=
2bs^2h.
\label{eq:av}
\end{equation}
Although GQA shares K/V across groups of query heads, the score and weighted-sum
operations are still performed per query head in the standard formulation, so
Eqs.~\eqref{eq:qk}--\eqref{eq:av} apply to both MHA and GQA.

\subsubsection{Output projection}

The concatenated attention output is projected with
$\mathbf{W}^O\in\mathbb{R}^{h\times h}$:
\begin{equation}
\mathrm{FLOPs}_{\mathrm{out}}
=
2bsh^2.
\label{eq:wo}
\end{equation}

\subsubsection{Total attention FLOPs per token}

For MHA,
\begin{align}
\mathrm{FLOPs}_{\mathrm{attn}}^{\mathrm{MHA}}
&=
6bsh^2 + 2bs^2h + 2bs^2h + 2bsh^2 \nonumber\\
&=
8bsh^2 + 4bs^2h,
\label{eq:attn_mha_total}
\end{align}
and therefore
\begin{equation}
\boxed{
\mathrm{FLOPs/token}_{\mathrm{attn}}^{\mathrm{MHA}}
=
8h^2 + 4sh.
}
\label{eq:attn_mha_token}
\end{equation}

For GQA,
\begin{align}
\mathrm{FLOPs}_{\mathrm{attn}}^{\mathrm{GQA}}
&=
2bsh^2\left(1+\frac{2}{r}\right)
+4bs^2h+2bsh^2 \nonumber\\
&=
4bsh^2+\frac{4bsh^2}{r}+4bs^2h,
\label{eq:attn_gqa_total}
\end{align}
and therefore
\begin{equation}
\boxed{
\mathrm{FLOPs/token}_{\mathrm{attn}}^{\mathrm{GQA}}
=
4h^2+\frac{4h^2}{r}+4sh.
}
\label{eq:attn_gqa_token}
\end{equation}

For a decoder-only Transformer with $L$ layers, the forward attention FLOPs per
token are $L$ times the corresponding per-layer quantity. For training, a
common approximation is forward plus backward $\approx 3\times$ forward FLOPs.

\subsubsection{What KV sharing removes under GQA}

Cross-layer KV sharing removes the new K/V projections for a shared layer.
Therefore, the forward projection FLOPs saved per shared layer and token are
\begin{equation}
\Delta F_{\mathrm{GQA}}
=
\frac{4h^2}{r}.
\label{eq:gqa_skv_flop_saving}
\end{equation}
If each cached element occupies $b_{\mathrm{elem}}$ bytes, the per-token
KV-cache reduction for a shared layer is
\begin{equation}
\Delta B_{\mathrm{GQA}}
=
2\frac{h}{r}b_{\mathrm{elem}}.
\label{eq:gqa_skv_cache_saving}
\end{equation}
Thus, GQA reduces the amount of K/V redundancy available to remove by the
grouping ratio $r$, but the savings remain non-zero.

\subsection{Inference Validation under GQA}
\label{app:gqa_validation}

We complement the FLOP analysis with decoding measurements on two GQA models.
Table~\ref{tab:app_gqa} shows that the inference-side benefit of KV sharing
persists even when K/V projections have already been reduced by grouped-query
attention.

\begin{table}[t]
\centering
\small
\caption{
\textbf{GQA inference validation.}
KV-Pipe improves decoding throughput on GQA-based models.
}
\label{tab:app_gqa}
\begin{tabular}{llcc}
\toprule
\textbf{Model} &
\textbf{Variant} &
\textbf{Throughput} &
\textbf{Time/token} \\
\midrule

LLaMA3-8B, 32K
& GQA baseline
& 620 tok/s
& 1.61 ms \\
& KV-Pipe, SKV=4
& \textbf{665 tok/s}
& \textbf{1.50 ms} \\

\midrule

Qwen2.5-14B, 128K
& GQA baseline
& 170 tok/s
& 5.88 ms \\
& KV-Pipe, SKV=6
& \textbf{184 tok/s}
& \textbf{5.43 ms} \\

\bottomrule
\end{tabular}
\end{table}

KV-Pipe improves throughput by approximately 7.3\% on LLaMA3-8B and 8.2\% on
Qwen2.5-14B. These results extend the inference case study beyond MHA and
support the ``double benefit'' claim under a more deployment-relevant attention
configuration, while the analysis above explains why the absolute K/V
projection saving is smaller than under MHA.

\subsection{Practical Operating Regime}
\label{app:practical_guidance}

The experiments identify a clear operating regime in which KV-Pipe is most
useful. The benefit grows when stage imbalance is large enough to materially
limit utilization. On Ascend 910B, the relative MFU improvement increases from
3.10\% at PP=2 to 4.32\% at PP=4 and 9.17\% at PP=8, while iteration-time
reduction increases from 4.90\% to 7.56\% and 9.80\%, respectively.
The GPU experiments in the main paper show the same qualitative trend across
model families and context lengths.

Accordingly, KV-Pipe is most attractive for moderate-to-high PP degrees,
long-context workloads, and partitions with non-uniform components such as a
heavy LM head or heterogeneous attention layers. At low PP degree or when the
baseline stage profile is already well balanced, the available headroom is
smaller and the architectural change may provide only a modest benefit.

\subsection{Discussion}
\label{sec:discussion}
KV-Pipe turns cross-layer KV sharing into a \emph{pipeline-level} optimization knob: by selectively converting full-attention layers into KV-sharing layers in FLOP-heavy pipeline stages, it reduces stage skew and improves end-to-end pipeline efficiency. Our Ascend 910B results (Table~\ref{tab:npu_best_subtables}, Figures~\ref{fig:npu_mfu_sweeps}--\ref{fig:npu_timeflops_sweeps}) show consistent gains across PP settings. Most notably, the relative MFU improvement grows with pipeline depth (PP): the baseline becomes more sensitive to stage imbalance and bubble overhead at larger PP, increasing the headroom for KV-Pipe. This is reflected both in higher MFU and reduced iteration time, indicating that KV-Pipe improves \emph{true} utilization rather than merely shifting work between phases.

A central observation from Figures~\ref{fig:npu_mfu_sweeps}--\ref{fig:npu_timeflops_sweeps} is that MFU is \emph{non-monotonic} in the shared-KV size.
Increasing KV sharing always reduces total model FLOPs, yet MFU increases only up to an optimal point and then declines. KV-Pipe explains this behavior through the proposed FLOPs Imbalance Ratio (FIR): small-to-moderate KV sharing reduces the bottleneck stage cost and drives FIR toward 1, minimizing idle time and bubbles. Past the optimum, KV-Pipe can over-correct the original bottleneck and shift the critical path to earlier stages, increasing FIR again and reintroducing bubbles even though total FLOPs keep decreasing. This highlights an important systems insight: for pipeline efficiency, the dominant objective is not minimizing FLOPs globally, but minimizing stage imbalance (i.e., aligning per-stage times).

Many PP techniques improve utilization primarily through scheduling (e.g., reducing bubbles via refined 1F1B variants, sequential schedules, interleaving, or micro-batch shaping). These methods can be highly effective, but they largely treat the per-stage compute profile as fixed and aim to hide bubbles. In contrast, KV-Pipe directly changes the per-stage workload by altering the effective cost of attention layers, thereby removing bubbles at the source. This distinction is important because stage imbalance often persists even with aggressive scheduling, especially in realistic models where the last stage contains extra modules (e.g., LM head) or where attention/MLP costs are non-uniform across depth.

KV-Pipe is intentionally designed to be orthogonal to pipeline scheduling and
parallel runtime choices. Conceptually, KV-Pipe outputs a modified per-layer
cost profile by introducing KV-sharing layers; it does not require changing the
pipeline scheduler or micro-batching policy. When a KV-sharing layer and its
source layer reside on different pipeline stages, the source K/V tensors must
additionally be transferred across the corresponding stage boundary. This
increases the communication volume at that boundary; however, the transfer can
be overlapped with pipeline computation and communication, so the additional
communication does not necessarily translate proportionally into end-to-end
latency. Therefore, KV-Pipe can be applied on top of a broad class of PP methods
as long as they preserve the same notion of \emph{stage-level execution} and the
primary bottleneck remains determined by the maximum stage time. The key
requirement for composability is that the PP method's efficiency is sensitive
to the distribution of compute across stages (i.e., the bottleneck stage
dominates). Under this condition, scheduling or partitioning methods can benefit
from KV-Pipe because it pushes the system toward a flatter stage-time profile
while allowing the additional cross-stage KV communication to be overlapped
with pipeline execution.

While KV-Pipe is motivated by training-time PP efficiency, it also inherits the classical advantage of KV-sharing for inference: reducing redundant KV projections and KV-cache growth. This yields a \emph{double benefit}: (i) improved training MFU by balancing stage workloads, and (ii) improved inference throughput by lowering per-token work and memory pressure, enabling larger effective batch sizes and faster decode. Importantly, these benefits arise from the same architectural lever (KV reuse) and do not require separate mechanisms for training and inference. As a result, KV-Pipe can be viewed as a \emph{native} system--architecture co-design knob: it improves PP efficiency during training and improves throughput/latency during serving, using a single consistent strategy grounded in stage-aware compute balancing.

\section*{Limitations and Future Work}

The quality results in this work should be interpreted within their measured
scope. We validate perplexity and downstream accuracy for LLaMA2-7B; the
LLaMA2-13B, LLaMA3-8B, and Qwen2.5-14B experiments provide system-level
efficiency evidence but are not used to claim architecture-independent quality
preservation. In particular, we do not assume that observations established for
dense LLaMA-family models automatically transfer to MoE, hybrid
Transformer--SSM architectures, or arbitrary training regimes.

Likewise, the current study evaluates PP degrees up to 8. Evaluation on
PP=16/32 and very large models remains future work. At larger scales the
relative contribution of the LM head and the location of the critical stage may
change, making the bottleneck-retargeting behavior more important. Finally,
the greedy Architecture-Balanced procedure is designed as a lightweight
offline heuristic; we do not claim global optimality over all possible
layer-sharing assignments. Joint optimization of pipeline placement,
KV-sharing placement, and an explicit quality constraint is an interesting
direction for future work.



\end{document}